\documentclass[twocolumn,prb,amsmath,showpacs,amssymb]{revtex4}
\usepackage{graphicx}% Include figure files
\usepackage{dcolumn}% Align table columns on decimal point
\usepackage{bm}% bold math

\begin{document}

\title{Dynamics and thermalization of the nuclear spin bath in the single-molecule magnet Mn$_{12}$-ac: test for the theory of spin tunneling.}
\author{Andrea Morello}
\affiliation{Kamerlingh Onnes Laboratory, Leiden University, P.O.
Box 9504, 2300RA Leiden, The Netherlands} \affiliation{Department of
Physics and Astronomy, University of British Columbia, Vancouver BC
V6T 1Z1, Canada} \affiliation{ARC Centre of Excellence for Quantum
Computer Technology, School of Electrical Engineering and
Telecommunications, The University of New South Wales, Sydney NSW
2052, Australia.} \email{a.morello@unsw.edu.au}

\author{L. J. de Jongh}
\affiliation{Kamerlingh Onnes Laboratory, Leiden University, P.O. Box 9504, 2300RA Leiden, The Netherlands.}

\date{\today}

\begin{abstract}
The description of the tunneling of a macroscopic variable in the
presence of a bath of localized spins is a subject of great
fundamental and practical interest, and is relevant for many
solid-state qubit designs. Most of the attention is usually given to
the dynamics of the ``central spin'' (i.e., the qubit), while little
is known about the spin bath itself. Here we present a detailed
study of the dynamics of the nuclear spin bath in the Mn$_{12}$-ac
single-molecule magnet, probed by NMR experiments down to very low
temperatures ($T \simeq 20$ mK). The results are critically analyzed
in the framework of the Prokof'ev-Stamp theory of nuclear-spin
mediated quantum tunneling. We find that the longitudinal relaxation
rate of the $^{55}$Mn nuclei in Mn$_{12}$-ac becomes roughly
$T$-independent below $T \simeq 0.8$ K, and can be strongly
suppressed with a longitudinal magnetic field. This is consistent
with the nuclear relaxation being caused by quantum tunneling of the
molecular spin, and we attribute the tunneling fluctuations to the
minority of fast-relaxing molecules present in the sample. The
transverse nuclear relaxation is also $T$-independent for $T < 0.8$
K, and can be explained qualitatively and quantitatively by the
dipolar coupling between like nuclei in neighboring molecules. This
intercluster nuclear spin diffusion mechanism is an essential
ingredient for the global relaxation of the nuclear spin bath. We
also show that the isotopic substitution of $^1$H by $^2$H leads to
a slower nuclear longitudinal relaxation, consistent with the
decreased tunneling probability of the molecular spin. Finally, we
demonstrate that, even at the lowest temperatures - where only
$T$-independent quantum tunneling fluctuations are present - the
nuclear spins remain in thermal equilibrium with the lattice
phonons, and we investigate the timescale for their thermal
equilibration. After a review of the theory of macroscopic spin
tunneling in the presence of a spin bath, we argue that most of our
experimental results are consistent with that theory, but the
thermalization of the nuclear spins is not. This calls for an
extension of the spin bath theory to include the effect of
spin-phonon couplings in the nuclear-spin mediated tunneling
process.
\end{abstract}

\pacs{75.45.+j, 76.60.-k, 03.65.Yz}% PACS, the Physics and Astronomy
                             % Classification Scheme.
%\keywords{Molecular magnet; nuclear relaxation; quantum tunneling; spin bath.}%Use showkeys class option if keyword
                              %display desired
\maketitle

\section{Introduction}

The understanding of quantum tunneling in mesoscopic systems has
made huge progress in the past decades, to the point that
nanofabricated devices are now being exploited as coherently
tunneling two-level systems (TLSs) for quantum information
purposes.\cite{nakamura99N,vion02S,chiorescu03S} Conceptually, a
first breakthrough was the proper description of the coupling of an
effective TLS to an environment described by an oscillator
bath.\cite{leggett87RMP} Whether the system is an intrinsic TLS
(e.g. a spin $s = 1/2$) or the low-energy truncation of a more
complicated entity (e.g. the flux state of a SQUID), one can
generally apply the oscillator bath theory when the environment is
described by delocalized modes (conduction electrons, phonons,
photons, etc.) and the couplings of the TLS to each oscillator are
weak. In many solid-states systems, however, it can be necessary to
account for localized environmental excitations whose couplings to
the TLS are not weak. This type of environment is called ``spin
bath'' \cite{prokof'ev95CM,prokof'ev96JLTP,prokof'ev00RPP} and
cannot be mapped onto an oscillator bath. Importantly, a spin bath
environment can cause decoherence even at $T=0$ and is therefore of
great relevance for quantum systems that are designed to show
coherent dynamics, like qubits for quantum computation. The
prototypical realization of a tunneling TLS coupled to a spin bath
is the giant spin of a single-molecule magnet
(SMM).\cite{gatteschi94S,christou00MRS,gatteschi03AC} These
molecular systems consist of a core of strongly interacting
transition metal ions, surrounded by organic ligands. At
sufficiently low temperatures the core of the molecule behaves
effectively like a single large spin $\mathbf{S}$. When uniaxial
magnetic anisotropy is present, the reversal of the spin direction
requires - classically - a large energy, so that the spin direction
can be frozen at very low $T$. However, in the presence of a
transverse magnetic field or a biaxial anisotropy, the spin
direction can be reversed by tunneling through the anisotropy
barrier.\cite{chudnovsky88PRL} The electronic spins that form the
SMM are magnetically coupled to the nuclear spins that either belong
to the magnetic ions themselves ($^{55}$Mn, $^{56}$Fe, \ldots) or to
the surrounding ligand molecules ($^1$H, $^{13}$C, \ldots). As a
consequence of these couplings, the observation of macroscopic
quantum tunneling of magnetization in SMMs
\cite{thomas96N,friedman96PRL,hernandez96EPL,gatteschi03AC,sangregorio97PRL}
cannot be understood without invoking the dynamics of the nuclear
spins themselves.\cite{prokof'ev96JLTP} The theoretical predictions
for the role of nuclear spins in the magnetization tunneling of SMMs
\cite{prokof'ev98PRL} have been verified by a series of experiments
on the Fe$_8$ compound.\cite{wernsdorfer99PRL,wernsdorfer00PRL} Most
remarkably, this material allows to change the isotopic composition
of the sample, both by strengthening ($^{56}$Fe $\rightarrow$
$^{57}$Fe substitution) and weakening ($^1$H $\rightarrow$ $^2$H
substitution) the hyperfine couplings, while leaving the electronic
structure of the SMMs unaffected. As predicted, the rate of quantum
relaxation of the magnetization was found to be directly related to
the nuclear isotopic composition of the
sample.\cite{wernsdorfer00PRL} More recently, the effect of isotopic
substitution has been observed in the low-$T$ electronic specific
heat of Fe$_8$ (Ref. \onlinecite{evangelisti05PRL}) and in the
dephasing time of coherent electron spin precession in
Cr$_7$Ni.\cite{ardavan07PRL} Nuclear spin effects were also invoked
in the interpretation of $\mu$SR data in isotropic
molecules,\cite{keren07PRL} and in an alternative description of the
short-term magnetic relaxation in SMMs.\cite{villain05EPJB} All
these works have analyzed the effect of the nuclei on the dynamics
of the ``central spin'', but a crucial aspect of the theory of the
spin bath is that the tunneling of the central system has
repercussion on the dynamics of the bath itself, so that the latter
cannot be simply regarded as an independent source of ``noise''.
Until now, the experiments to probe the electron spin dynamics have
not been able to test this delicate aspect of the theory. To
understand the details of the nuclear spin fluctuations, one should
then look \emph{directly} at the nuclear spins by means of
low-temperature NMR experiments, performed under different regimes
for the quantum dynamics of the electron spin. These experiments
have been carried out by several
groups,\cite{morello04PRL,goto03PRB,ueda02PRB,baek05PRB,chakov06JACS}
but an accurate analysis of their implications for the more general
theory of nuclear-spin mediated quantum tunneling is still lacking.

In this work, we present a comprehensive set of experiments on the
dynamics of $^{55}$Mn nuclear spins in the Mn$_{12}$-ac SMM, and we
use our results for a critical assessment of the theory of the spin
bath. Our data provide definitive proof that the nuclear spin
dynamics is strongly correlated with that of the central spin, that
is, it cannot be treated as an independent source of noise. Indeed,
we find that the nuclear spin fluctuations change dramatically when
the tunneling dynamics of the central spin is modified, e.g. by an
external magnetic field. In addition, we shall demonstrate that the
nuclear spins remain in thermal equilibrium with the phonon bath
down to the lowest temperatures ($T \simeq 20$ mK) accessible to our
experiment, where the thermal fluctuations of the electron spins are
entirely frozen out. This implies that there is a mechanism for
exchanging energy between nuclei, electrons and phonons
\emph{through the nuclear-spin mediated quantum tunneling of the
central spin}. This is the point where the current theoretical
description of macroscopic quantum tunneling in the presence of a
spin bath needs to be improved.

As regards the ``macroscopicness'' of the quantum effects observed in SMMs, we adopt Leggett's view that the most
stringent criterion is the ``disconnectivity'',\cite{leggett80SPTP,leggett02JPCM} $\mathcal{D}$, which roughly
speaking is the number of particles that behave differently in the two branches of a quantum superposition. For
instance, while a Cooper pair box\cite{nakamura99N} is a relatively large, lithographically fabricated device,
the quantum superposition of its charge states involves in fact only one Cooper pair, i.e. two electrons, and its
disconnectivity is only $\mathcal{D}=2$. The matter-wave interference in fullerene molecules,\cite{arndt99N} for
instance, is a much more ``quantum macroscopic'' phenomenon, since it means that 60 $\times$ (12 nucleons + 6
electrons) = 1080 particles are superimposed between different paths through a diffraction grating. For the spin
tunneling in Mn$_{12}$-ac SMMs discussed here, we have 44 electron spins simultaneously tunneling between
opposite directions, which places this system logarithmically halfway between single particles and fullerenes on
a macroscopicness scale.

The paper is organized as follows. Section \ref{Experimental} describes the physical properties of the sample
used in the experiments, the design and performance of our measurement apparatus, and the methods of data
analysis. Section \ref{nsd} presents the experimental results on the nuclear spin dynamics, starting with the NMR
spectra, the longitudinal and transverse relaxation rates in zero field, and their dependence on a longitudinal
external field. We also study the nuclear relaxation in different Mn sites within the cluster, and the effect of
isotopic substitution in the ligand molecules. In Section \ref{spinT} we discuss the thermal equilibrium between
nuclear spins and phonon bath, the experimental challenges in optimizing it, and the indirect observation of
magnetic avalanches during field sweeps. In Section \ref{theory} we give an introductory review of the theory of
the spin bath, and apply its predictions to the calculation of the nuclear relaxation rate as observed in our
experiments. Together with the information on the thermal equilibrium of the nuclear spins, this will allow us to
draw clear-cut conclusions on the status of our current theoretical understanding of quantum tunneling of
magnetization. We conclude with a summary and implications of the results in Section \ref{conclusions}.

\section{Experiment} \label{Experimental}

\subsection{Sample properties} \label{sample}

\begin{figure*}[t]
\includegraphics[width=12cm]{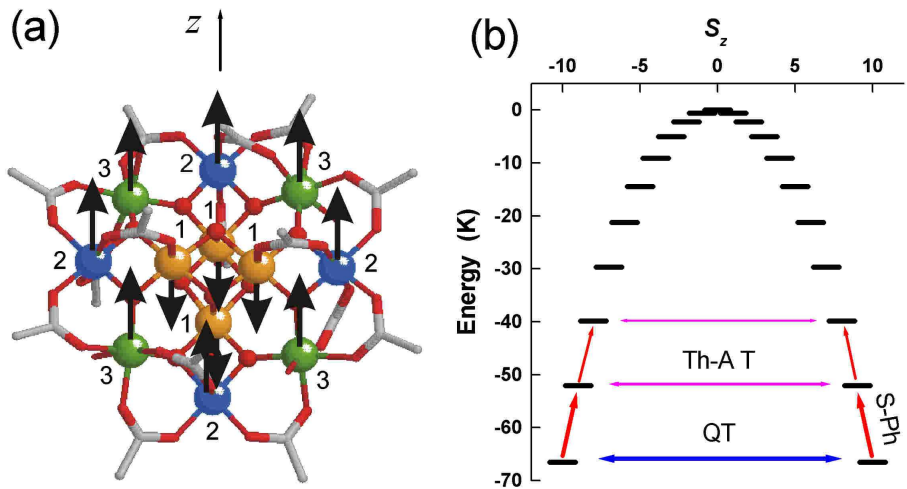}
\caption{\label{structure} (Color online) (a) Structure of the Mn$_{12}$-ac cluster, with the labelling of the
three inequivalent Mn sites as described in the text. (b) Energy level scheme for the electron spin as obtained
from the Hamiltonian (\ref{hamiltonianMn12}), retaining only the terms diagonal in $S_z$. The non-diagonal terms
allow transitions between states on opposite sides of the anisotropy barrier by means of quantum tunneling (QT).
In the presence of intrawell transitions induced by spin-phonon interaction (S-Ph), thermally assisted quantum
tunneling (Th-A T) between excited doublets can also take place.}
\end{figure*}

We chose to focus our study on the well-known
[Mn$_{12}$O$_{12}$(O$_{2}$CMe)$_{16}$(H$_{2}$O)$_{4}$]
(Mn$_{12}$-ac) compound, which belongs to the family of SMMs with
the highest anisotropy barrier. As we shall see below, the rationale
for choosing a SMM with high anisotropy barrier is that the electron
spin fluctuations become slow on the NMR timescale already at
temperatures of a few kelvin. The structure of the cluster
\cite{lis80AC} (Fig. \ref{structure}) consists of a core of 4
Mn$^{4+}$ ions with electron spin $s = 3/2$, which we shall denote
as Mn$^{(1)}$, and 8 Mn$^{3+}$ ions ($s = 2$) on two inequivalent
crystallographic sites, Mn$^{(2)}$ and Mn$^{(3)}$ [Fig.
\ref{structure}(a)]. Within the molecular cluster, the electron
spins are coupled by mutual superexchange interactions, the
strongest being the antiferromagnetic interaction between Mn$^{(1)}$
and Mn$^{(2)}$ (Ref. \onlinecite{sessoli93JACS}). The molecules
crystallize in a tetragonal structure with lattice parameters
$a=b=17.319$ \AA~ and $c=12.388$ \AA. The ground state of the
molecule has a total electron spin $S=10$ and, for the temperature
range of interest in the present work ($T<2$ K), we may describe the
electron spin of the cluster by means of the effective spin
Hamiltonian:
\begin{eqnarray}
\mathcal{H}=-DS_z^2 - BS_z^4 + E(S_x^2 - S_y^2) - C(S_+^4 + S_-^4) + \nonumber \\
+ \mu_B \mathbf{B} \cdot \mathbf{g} \cdot \mathbf{S}. \label{hamiltonianMn12}
\end{eqnarray}
Commonly adopted parameter values are $D = 0.548$ K, $B = 1.17$ mK and $C = 22$ $\mu$K as obtained by neutron
scattering data,\cite{mirebeau99PRL} and for the $\mathbf{g}$ tensor the values $g_{\parallel} = 1.93$ and
$g_{\perp} = 1.96$ from high-frequency EPR.\cite{barra97PRB,hill98PRL,noteanisotropy} The uniaxial anisotropy
terms $-DS_z^2$ and $- BS_z^4$ can be attributed to the single-ion anisotropy of the Mn$^{3+}$
ions,\cite{barra97PRB} which is due to the crystal field effects resulting in the Jahn-Teller distortions of the
coordination octahedra, where the elongation axes are approximately parallel to the $\hat{c}$-axis of the
crystal. Considering only the diagonal terms, the energy levels scheme would be a series of doublets of
degenerate states, $|\pm m \rangle$, separated by a barrier with a total height $DS^2 + BS^4 \simeq 66.6$ K [Fig.
\ref{structure}(b)]. The transverse anisotropy terms, $E(S_x^2 - S_y^2) - C(S_+^4 + S_-^4)$, lift the degeneracy
of the $|\pm m\rangle$ states and allow quantum tunneling of the giant spin through the anisotropy barrier. We
call $\Delta_m$ the matrix element for the tunneling of the giant spin through the $m$-th doublet, and
$2\Delta_m$ the corresponding tunneling splitting. The $C(S_+^4 + S_-^4)$ term arises from the fourfold $S_4$
point symmetry of the molecule, but there is now solid experimental evidence \cite{hill03PRL,delbarco03PRL} for
the prediction \cite{cornia02PRL} that a disorder in the acetic acid of crystallization is present and gives rise
to six different isomers of Mn$_{12}$ cluster, four of which have symmetry lower than tetragonal and therefore
have nonzero rhombic term $E(S_x^2 - S_y^2)$. EPR experiments give an upper bound $E \leq 14$ mK.\cite{hill03PRL}
For the purpose of NMR experiments, such isomerism may cause slight variations in the local hyperfine couplings,
causing extra broadening in the $^{55}$Mn resonance lines. Very recently, a new family of Mn$_{12}$ clusters has
been synthesized, which does not suffer from the solvent disorder mentioned above, and yields indeed more sharply
defined $^{55}$Mn NMR spectra.\cite{harter05IC}

When adding spin-phonon
interactions,\cite{hartmann96IJMPB,leuenberger00PRB} the possible
transitions between the energy levels of (\ref{hamiltonianMn12}) are
sketched in Fig. \ref{structure}(b). We distinguish between
\emph{intrawell} spin-phonon excitations, where the spin state
remains inside the same energy potential well, and the
\emph{interwell} transitions, which involve spin reversal by quantum
tunneling through the barrier, allowed by the terms in
(\ref{hamiltonianMn12}) that do not commute with $S_z$.
Thermally-assisted tunneling involves both these types of
transitions.

The above discussion refers to the majority of the molecules in a
real sample, but for our experiments the crucial feature of
Mn$_{12}$-ac is the presence of fast-relaxing molecules
(FRMs),\cite{aubin97CC} i.e. clusters characterized by a lower
anisotropy barrier and a much faster relaxation rate, as observed
for instance by \textit{ac}-susceptibility\cite{evangelisti99SSC}
and magnetization measurements.\cite{wernsdorfer99EPL} It has been
recognized that such FRMs originate from Jahn-Teller
isomerism,\cite{sun99CC} i.e. the presence in the molecule of one or
two Mn$^{3+}$ sites where the elongated Jahn-Teller axis points in a
direction roughly perpendicular instead of parallel to the
crystalline $\hat{c}$-axis. This results in the reduction of the
anisotropy barrier to 35 or 15 K in the case of one or two flipped
Jahn-Teller axes, respectively,\cite{wernsdorferU} and presumably in
an increased strength of the non-diagonal terms in the spin
Hamiltonian as well. Furthermore, the anisotropy axis $z$ of the
whole molecule no longer coincides with the crystallographic
$\hat{c}$-axis, but deviates e.g. by $\sim 10^{\circ}$ in the
molecules with 35 K barrier.\cite{wernsdorfer99EPL} The Jahn-Teller
isomerism is very different from the above-mentioned effect of
disorder in solvent molecules, and produces much more important
effects for the present study. As will be argued below, the presence
of the FRMs is essential for the interpretation of our results and,
to some extent, may be regarded as a fortunate feature for this
specific experiment.

The sample used in the experiment consisted of about 60 mg of polycrystalline Mn$_{12}$-ac, with typical
crystallite volume $\sim 0.1$ mm$^3$. The crystallites were used as-grown (i.e., not crushed), mixed with Stycast
1266 epoxy, inserted in a $\varnothing$ 6 mm capsule and allowed to set for 24 hours in the room temperature bore
of a 9.4 T superconducting magnet. With this procedure, the magnetic easy axis of the molecules (which coincides
with the long axis of the needle-like crystallites) ends up being aligned along the field within a few degrees.
In addition, we shall report NMR spectra taken on a small single crystal (mass $\sim 1$ mg).

\subsection{Low-temperature pulse NMR setup}

\begin{figure*}[t]
\includegraphics[width=14.5cm]{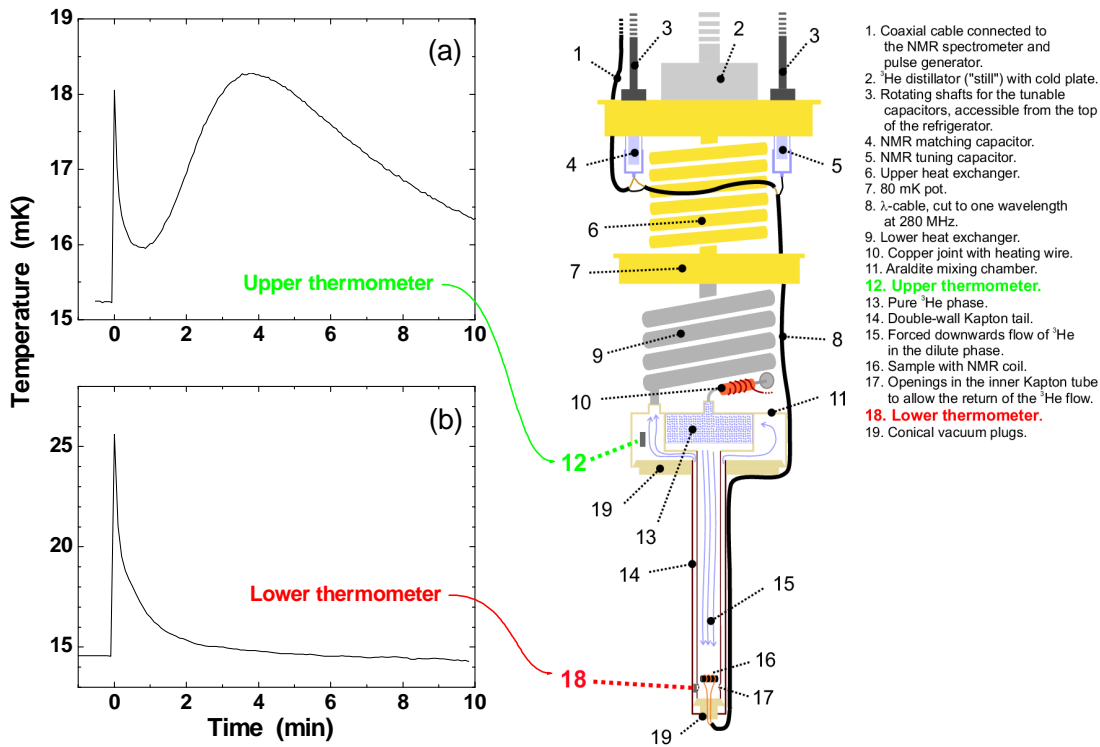}
\caption{(Color online) Sketch of the low-temperature part of the
dilution refrigerator, showing the components of the NMR circuitry,
the special plastic mixing chamber and the position of the
thermometers. Graph panels: temperatures recorded at the (a) upper
and (b) lower mixing chamber thermometers, having applied a
spin-echo NMR pulse sequence at time $t=0$.}
\label{dilution+temperatures}
\end{figure*}

Our experimental setup is based on a Leiden Cryogenics MNK126-400ROF dilution refrigerator, fitted with a plastic
mixing chamber that allows the sample to be thermalized directly by the $^3$He flow. A scheme of the
low-temperature part of the refrigerator is shown in Fig. \ref{dilution+temperatures}, together with the NMR
circuitry. The mixing chamber consists of two concentric tubes, obtained by rolling a Kapton foil coated with
Stycast 1266 epoxy. The tops of each tube are glued into concentric Araldite pots: the inner pot receives the
downwards flow of condensed $^3$He and, a few millimeters below the inlet, the phase separation between the pure
$^3$He phase and the dilute $^3$He/$^4$He phase takes place. The circulation of $^3$He is then forced downwards
along the inner Kapton tube, which has openings at the bottom side to allow the return of the $^3$He stream
through the thin space in between the tubes. Both the bottom of the Kapton tail and the outer pot are closed by
conical Araldite plugs smeared with Apiezon N grease.

A 2-turns copper coil is wound around the capsule containing the sample, mounted on top of the lower conical plug
and inserted in the $^3$He/$^4$He mixture at the bottom of the mixing chamber tail, which coincides with the
center of a 9 T superconducting magnet. The coil is then connected by a thin brass coaxial cable (length $\approx
0.5$ m) to two tunable cylindrical teflon capacitors, mounted at the still (see Fig.
\ref{dilution+temperatures}). At the frequency where the cable connecting capacitors and coil is precisely one
wavelength, the circuit is equivalent to a standard lumped $LC$-resonator. However, since the $\lambda$-cable is
a low-conductivity coax for low-$T$ applications, the quality factor of the resonator (which includes the cable)
is drastically reduced. Although this affects the sensitivity of the circuit, it also broadens the accessible
frequency range without the need to retune the capacitors. Cutting the cable for one wavelength at $\sim 280$
MHz, the circuit is usable between (at least) 220 and 320 MHz. As for the room-temperature NMR electronics,
details can be found in Ref. \onlinecite{morelloT}.

The temperature inside the mixing chamber is monitored by two
simultaneously calibrated Speer carbon thermometers, one in the
outer top Araldite pot, and the other at the bottom of the Kapton
tail, next to the sample. At steady state and in the absence of NMR
pulses, the temperature along the mixing chamber is uniform within
$\lesssim 0.5$ mK. The effect of applying high-power ($\sim 100$ W)
NMR pulses is shown in Fig. \ref{dilution+temperatures}(a) and (b).
A sudden increase in the measured temperature is seen both at the
bottom and the top thermometer, and can be attributed to the short
electromagnetic pulse. The temperature at the lower thermometer,
i.e. next to the sample and the NMR coil, quickly recovers its
unperturbed value, whereas the upper thermometer begins to sense the
``heat wave'' carried by the $^3$He stream with a delay of about 3
minutes. This has the important consequence that we can use the
upper thermometer to distinguish the effect of sudden
electromagnetic radiation bursts from the simple heating of the
$^3$He/$^4$He mixture, as will be shown in \S\ref{superradiance}
below.

The sample temperature is regulated by applying current to a
manganin wire, anti-inductively wound around a copper joint just
above the $^3$He inlet in the mixing chamber. In this way we can
heat the incoming $^3$He stream and uniformly increase the mixing
chamber temperature.

For the $^3$He circulation we employ an oil-free pumping system, consisting of a 500 m$^3$/h Roots booster pump,
backed by two 10 m$^3$/h dry scroll pumps. The system reaches a base temperature of 9 mK, and the practical
operating temperature while applying $rf$-pulses is as low as 15 - 20 mK.

\subsection{Measurements and data analysis} \label{measures}

The $^{55}$Mn nuclear precession was detected by the spin-echo technique. A typical pulse sequence includes a
first $\pi / 2$-pulse with duration $t_{\pi / 2} = 12$ $\mu$s, a waiting interval of 45 $\mu$s, and a 24 $\mu$s
$\pi$-pulse for refocusing. Given the heating effects shown in Fig. \ref{dilution+temperatures}, a waiting time
of 600 s between subsequent pulse trains easily allows to keep the operating temperature around $15-20$ mK.
Moreover, at such low temperature the signal intensity is so high that we could obtain an excellent
signal-to-noise ratio without need of averaging, so that a typical measurement sequence took less than 12 hours.
Above 100 mK it proved convenient to take a few averages, but there the heating due to the \textit{rf}-pulses
became negligible, and the waiting time could be reduced to $\sim 100$ s.

\begin{figure}[t]
\includegraphics[width=8cm]{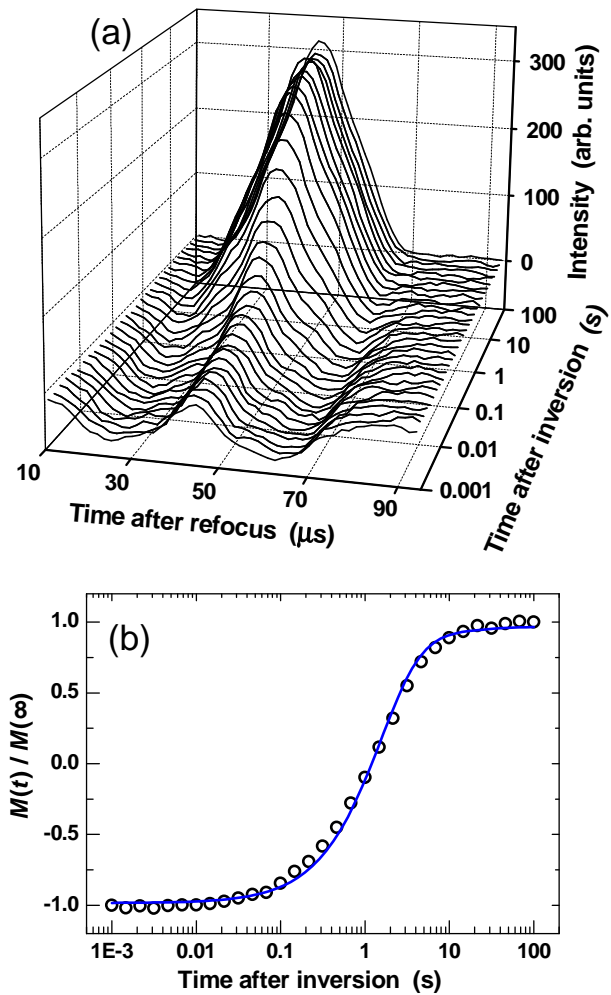}
\caption{(Color online) (a) An example of ``real time'' echo signals recorded during an inversion recovery, i.e.
measuring the echo intensity at increasing delays after an inversion pulse. In particular, these are single-shot
(no averaging) raw data taken at $B=0$ and $T=20$ mK in the Mn$^{(1)}$ site. (b) The (normalized) integral of the
echoes (open dots) is fitted to Eq. (\ref{recovery}) (solid line) to yield the LSR rate $W$.}\label{echo3D}
\end{figure}

The longitudinal spin relaxation (LSR) was studied by measuring the
recovery of the longitudinal nuclear magnetization after an
inversion pulse. We preferred this technique to the more widely used
saturation recovery \cite{furukawa01PRB,kubo02PRB,goto03PRB} because
it avoids the heating effects of the saturation pulse train, but we
checked at intermediate temperatures that the two methods indeed
lead to the same value of LSR rate. An example of echo signals
obtained as a function of the waiting time after the inversion pulse
is shown in Fig. \ref{echo3D}(a). By integrating the echo intensity
we obtain the time-dependence of the nuclear magnetization, $M(t)$,
as shown in Fig. \ref{echo3D}(b). For the ease of comparison between
different curves, we renormalize the vertical scale such that
$M(0)/M(\infty)=-1$ and $M(t \gg T_1)/M(\infty)=1$, even though
usually $|M(0)|<|M(\infty)|$, as could be deduced from Fig.
\ref{echo3D}(a). This is just an artifact that occurs when the NMR
line is much broader than the spectrum of the inversion pulse, and
does not mean that the length of the $\pi$-pulse is incorrect. Since
the $^{55}$Mn nuclei have spin $I = 5/2$, we fitted the recovery of
the nuclear magnetization with: \cite{suter98JPCM}
\begin{eqnarray}
    \frac{M(t)}{M(\infty)} = 1 - \left[ \frac{100}{63} e^{-30 W t} + \frac{16}{45} e^{-12 W t}
    + \frac{2}{35} e^{-2 W t} \right]
\label{recovery}
\end{eqnarray}
where $W$ is the longitudinal spin relaxation rate. Note that, in the simple case of a spin 1/2, $W$ is related
to the relaxation time $T_1$ by $2W=T_1^{-1}$. The above multiexponential expression and its numerical
coefficients are derived under the assumption that the $I=5/2$ multiplet is split by quadrupolar interactions,
and it is possible to resolve the central transition within that multiplet. While earlier work indicated that all
three manganese NMR lines are quadrupolar-split,\cite{kubo02PRB} more recent experiments on single crystal
samples have questioned that conclusion,\cite{harter05IC,chakov06JACS} and thereby the applicability of Eq.
(\ref{recovery}) to the present experiments. Even if other sources of line broadening hinder the visibility of
the quadrupolar contribution, the condition for the absence of quadrupolar splitting is an exactly cubic
environment for the nuclear site, which is not satisfied here. For this reason, and for the ease of comparison
with our\cite{morello03POLY,morello04PRL} and other groups' earlier
results,\cite{furukawa01PRB,kubo02PRB,goto03PRB} we choose to retain Eq. (\ref{recovery}) for the analysis of the
inversion recovery data.

The transverse spin relaxation (TSR) rate $T_2^{-1}$ was obtained by measuring the decay of echo intensity upon
increasing the waiting time $\tau$ between the $\pi/2$- and the $\pi$-pulses. The decay of transverse
magnetization $M_{\perp}(\tau)$ can be fitted by a single exponential
\begin{eqnarray}
\frac{M_{\perp}(2\tau)}{M_{\perp}(0)}=\exp\left(-\frac{2\tau}{T_2}\right)
\label{T2}
\end{eqnarray}
except at the lowest temperatures ($T \lesssim 0.2$ K), where also
a gaussian component $T_{2G}^{-1}$ needs to be included:
\begin{eqnarray}
\frac{M_{\perp}(2\tau)}{M_{\perp}(0)}=\exp \left(
-\frac{2\tau}{T_{2}}\right)  \exp\left(
-\frac{(2\tau)^2}{2T_{2G}^2}\right) \label{T2LG}
\end{eqnarray}

As regards the experiments to determine the nuclear spin
temperature, the measurements were performed by monitoring the
echo intensity at regular intervals while changing the temperature
$T_{\mathrm{bath}}$ of the $^3$He/$^4$He bath in which the sample
is immersed. Recalling that the nuclear magnetization is related
to the nuclear spin temperature $T_{\mathrm{nucl}}$ by the Curie
law:
\begin{eqnarray}
M(T_{\mathrm{nucl}})=N \mu_0 \frac{\hbar^2 \gamma_N^2
I(I+1)}{3k_{\mathrm{B}}T_{\mathrm{nucl}}}, \label{curie}
\end{eqnarray}
and assuming that $T_{\mathrm{bath}} = T_{\mathrm{nucl}}$ at a
certain temperature $T_0$ (e.g. 0.8 K), we can define a
calibration factor $K$ such that $M(T_0) =
K/T_{\mathrm{nucl}}(T_0)$ and use that definition to derive the
time evolution of the nuclear spin temperature as
$T_{\mathrm{nucl}}(t) = K/M(t)$ while the bath temperature is
changed.

Due to the strong magnetic hysteresis of Mn$_{12}$-ac, it is important to specify the magnetization state of the
sample since, as will be shown below, this parameter can influence the observed nuclear spin dynamics. Therefore
we carried out experiments under both zero-field cooled (ZFC) and field-cooled (FC) conditions, which correspond
to zero and saturated magnetization along the easy axis, respectively. Heating the sample up to $T \approx 4$ K
is sufficient to wash out any memory of the previous magnetic state. When the sample is already at $T \ll 1$ K,
the field-cooling procedure can be replaced by the application of a longitudinal field large enough to destroy
the anisotropy barrier, e.g. $B_z = 8$ T. Importantly, the shift of the $^{55}$Mn NMR frequency with external
field depends on the magnetization state of the sample:\cite{kubo01PhyB,kubo02PRB} in a ZFC sample each resonance
line splits in two, one line moving to $\omega_0 + \gamma_{\rm N} B_z$ and the other to $\omega_0 - \gamma_{\rm
N} B_z$. Conversely, in a FC sample only one line is observed, shifting to higher or lower frequency depending on
the direction of $B_z$ relative to the magnetization direction. Therefore, by measuring the intensity of the
shifted lines in a moderate longitudinal field, typically $\sim 0.5$ T, we can check the magnetization of the
sample \emph{as seen by the nuclei that contribute to the NMR signal}.

\section{Nuclear spin dynamics} \label{nsd}

\subsection{NMR spectra}

\begin{figure}[t]
\includegraphics[width=8.5cm]{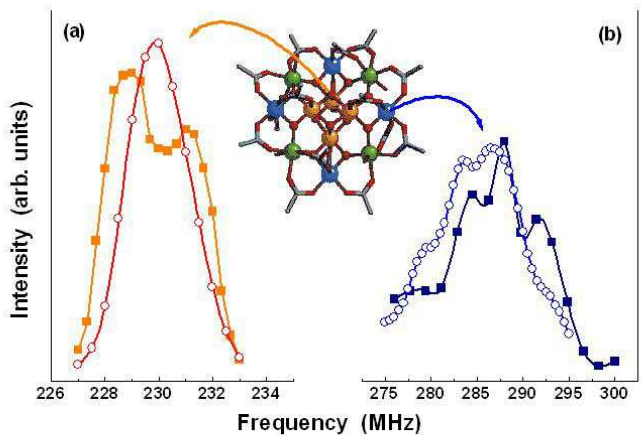}
\caption{(Color online) $^{55}$Mn NMR spectra of the (a) Mn$^{(1)}$
and (b) Mn$^{(2)}$ lines in Mn$_{12}$-ac, at $T=20$ mK. Open
circles: oriented powder. Solid squares: single crystal. The Mn
sites corresponding to each line are shown in the central drawing of
the molecular structure. All the spectra are measured in a
field-cooled sample.}\label{spectra}
\end{figure}

The basic feature of the $^{55}$Mn NMR spectra in Mn$_{12}$-ac is the presence of three well-separated lines,
that can be ascribed to three crystallographically inequivalent Mn sites in the molecule. The Mn$^{(1)}$ line,
centered around $\nu_1 \approx 230$ MHz, originates from the nuclei that belong to the central core of Mn$^{4+}$
ions, whereas the Mn$^{(2)}$ and Mn$^{(3)}$ lines, centered at $\nu_2 \approx 280$ and $\nu_3 \approx 365$ MHz,
respectively, have been assigned to the nuclei in the outer crown of Mn$^{3+}$
ions.\cite{furukawa01PRB,kubo02PRB} In Fig. \ref{spectra} we show the Mn$^{(1)}$ and Mn$^{(2)}$ spectra at $T=20$
mK, both in the oriented powder and in the single crystal, in a FC sample. Note that, whereas single-crystal
spectra of Mn$_{12}$-ac have been recently published,\cite{harter05IC} the present spectra are the only ones
measured at subkelvin temperatures so far. As argued already in Ref. \onlinecite{harter05IC}, the single-crystal
spectra indicate that the width of the Mn$^{(1)}$ line may not originate from a small quadrupolar splitting.
Instead, at least two inequivalent Mn$^{4+}$ sites may exist, supporting the growing amount of evidence about the
lack of symmetry of the Mn$_{12}$-ac compound.

We also note that the highest peak in the Mn$^{(2)}$ line at
$T=20$ mK is found at a frequency $\nu_2 \approx 287$ MHz about 8
MHz higher than most of the previously reported spectra at $T>1$
K,\cite{furukawa01PRB,kubo02PRB,harter05IC} with the exception of
Ref. \onlinecite{goto00phyB}, whereas the position of the
Mn$^{(1)}$ line is consistent with all the previous reports.

\subsection{Longitudinal spin relaxation in zero field} \label{LSRzerofield}

\begin{figure}[t]
\includegraphics[width=8.5cm]{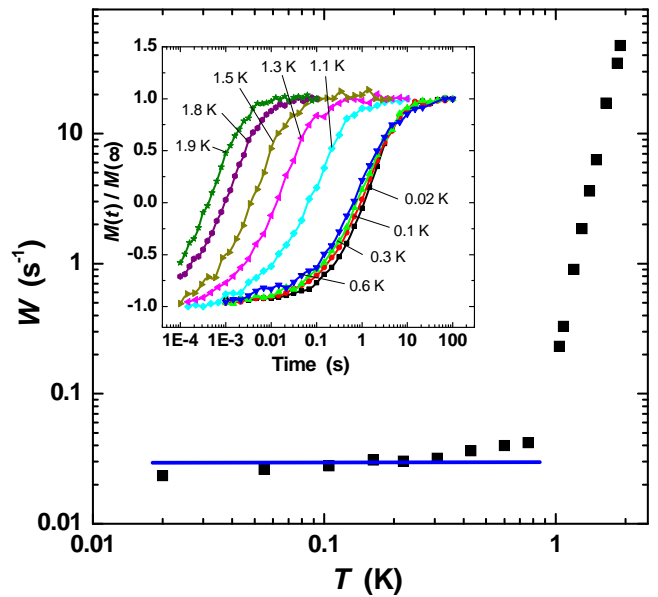}
\caption{(Color online) Temperature-dependence of the nuclear
spin-lattice relaxation rate $W$ of the Mn$^{(1)}$ line, in zero
external field and ZFC sample. The inset shows some examples of
recovery of the nuclear magnetization after a time $t$ from an
inversion pulse, at the indicated temperatures. These curves have
been fitted to Eq. (\ref{recovery}) to extract $W$.\label{Wzf}}
\end{figure}

The LSR rate as a function of temperature for the Mn$^{(1)}$ line,
in zero field and zero-field cooled (ZFC) sample, is shown in Fig.
\ref{Wzf}. The most prominent feature in these data is a sharp
crossover at $T \simeq 0.8$ K between a roughly exponential
$T$-dependence and an almost $T$-independent plateau. We have
previously attributed the $T$-independent nuclear relaxation to the
effect of tunneling fluctuations within the ground doublet of the
cluster spins,\cite{morello04PRL} and we shall dedicate most of the
present paper to discuss our further results supporting this
statement. Here we shall also argue that, even in the
high-temperature regime, thermally assisted quantum tunneling plays
an essential role, and the experimental results cannot be understood
simply in terms of LSR driven by intrawell electronic
transitions.\cite{furukawa01PRB} It should be noted that the
crossover from thermally activated to ground-state tunneling has
also been observed by analyzing the $T$-dependence of the steps in
the magnetization hysteresis
loops.\cite{chiorescu00PRL,bokacheva00PRL} The important advantage
of our NMR measurements is that the nuclear dynamics is sensitive to
\emph{fluctuations} of the cluster electron spins without even
requiring a change in the macroscopic magnetization of the sample.
Clearly, no macroscopic probe (except perhaps an extremely sensitive
magnetic noise detector) would be able to detect the presence of
tunneling fluctuations in a zero-field cooled sample in zero
external field, since the total magnetization is zero and remains
so. Below $T \sim 1.5$ K the steps in the hysteresis loops of
Mn$_{12}$-ac can be observed only at relatively high values of
external field,\cite{chiorescu00PRL,bokacheva00PRL} which means that
the spin Hamiltonian under those conditions is radically different
from the zero-field case. Therefore, that both our data and the
previous magnetization measurements show a crossover around $T
\simeq 0.8$ K should be considered as a coincidence.

The roughly $T$-independent plateau in the LSR rate below $T \simeq
0.8$ K is characterized by a value of $W \simeq 0.03$ s$^{-1}$ which
is surprisingly high, which at first sight may appear like an
argument against the interpretation in terms of tunneling
fluctuations of the electron spin. Experimentally it is indeed well
known\cite{thomas99PRL} that the relaxation of the magnetization in
Mn$_{12}$-ac in zero field may take years at low $T$, which means
that the tunneling events are in fact extremely rare. Based on this,
we are forced to assume that tunneling takes place only in a small
minority of the clusters, and that some additional mechanism takes
care of the relaxation of the nuclei in molecules that do not
tunnel. This is a very realistic assumption, since all samples of
Mn$_{12}$-ac are reported to contain a fraction of
FRMs,\cite{sun99CC,wernsdorfer99EPL} as mentioned in Sect.
\ref{sample}. Moreover, since we are also able to monitor the sample
magnetization, we verified that e.g. a FC sample maintains indeed
its saturation magnetization for several weeks \emph{while nuclear
relaxation experiments are being performed (at zero field)}. This
confirms that any relevant tunneling dynamics must originate from a
small minority of molecules. On the other hand, it also means that
the observed NMR signal comes mainly from nuclei belonging to frozen
molecules, thus there must be some way for the fluctuations in FRMs
to influence the nuclear dynamics in the majority of slow molecules
as well. One possibility is to ascribe it to the fluctuating dipolar
field produced by a tunneling FRM at the nuclear sites of
neighboring frozen molecules. In that case we may give an estimate
of $W$ using an expression of the form:
\begin{eqnarray}
W \approx \frac{\gamma_{\rm N}^2}{4} b_{\mathrm{dip}}^2
\frac{\tau_{\mathrm{T}}}{1 + \omega_{\rm N}^2 \tau_{\mathrm{T}}^2}
\approx
\frac{b_{\mathrm{dip}}^2}{4B_{\mathrm{tot}}}\tau_{\mathrm{T}}^{-1},
\label{Wdipolar}
\end{eqnarray}
where $b_{\mathrm{dip}}$ is the perpendicular component of the fluctuating dipolar field produced by a tunneling
molecule on its neighbors and $\tau_{\mathrm{T}}^{-1}$ is the tunneling rate. The highest value that
$b_{\mathrm{dip}}$ may take is $\sim 3$ mT in the case of nearest neighbors, which leads to the condition $W
\simeq 0.03$ s$^{-1}$ $\Rightarrow \tau_{\mathrm{T}}^{-1} \gg 10^6$ s$^{-1}$. Such a high rate is of course
completely unrealistic. We must therefore consider the effect of a tunneling molecule on the nuclei that
\emph{belong} to the molecule itself, and look for some additional mechanism that links nuclei in FRMs with
equivalent nuclei in frozen clusters. It is natural to seek the origin of such a mechanism in the intercluster
nuclear spin diffusion, and in the next section we shall provide strong experimental evidences to support this
interpretation.

\subsection{Transverse spin relaxation}

\begin{figure}[t]
\includegraphics[width=8.5cm]{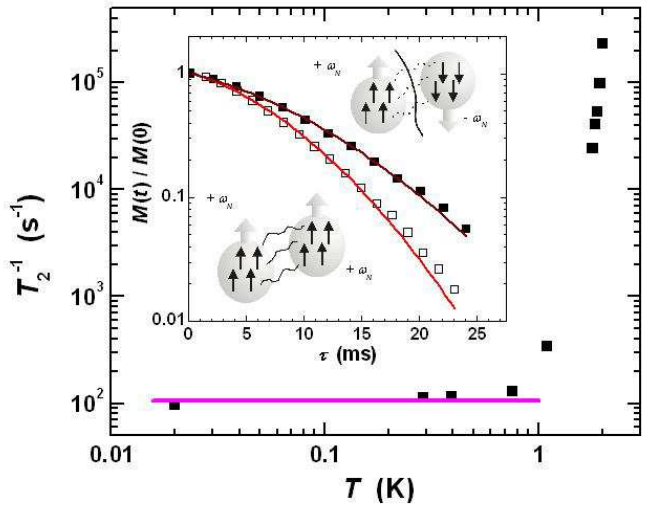}
\caption{\label{T2FCZFC} (Color online) Temperature-dependence of
the TSR rate $T_2^{-1}$ (squares) rates for a ZFC sample in zero
field and $\nu = 231$ MHz. The solid line in the $T$-independent
regime is a guide for the eye. Inset: normalized decay of transverse
nuclear magnetization, $M(\tau)/M(0)$, for ZFC (full squares) and FC
(open squares) sample, at $T = 20$ mK. The solid lines are fits to
Eq. (\ref{T2LG}), yielding the ratio
$T_{2G}^{-1}(\mathrm{FC})/T_{2G}^{-1}(\mathrm{ZFC}) = 1.35 \simeq
\sqrt{2}$. The sketches in the inset represent pictorially the fact
that intercluster spin diffusion is possible in a FC sample since
all the nuclei have the same Larmor frequency, contrary to the case
of a ZFC sample.}
\end{figure}

The $T$-dependence of the TSR rate $T_2^{-1}(T)$ is shown in Fig.
\ref{T2FCZFC}. One may observe that below 0.8 K the TSR, just like
the LSR, saturates to a nearly $T$-independent plateau. In
particular, $T_2^{-1}(T<0.8 \quad \mathrm{K}) \approx 100$ s$^{-1}$,
which is a factor $\sim 3000$ larger than the low-$T$ limit of the
LSR rate $W$. The values plotted in Fig. \ref{T2FCZFC} are all
obtained by fitting the decay of the transverse magnetization with
Eq. (\ref{T2}), i.e. with a single exponential. While this is very
accurate at high $T$, we found that for $T \lesssim 0.2$ K a better
fit is obtained by including a Gaussian component, as in Eq.
(\ref{T2LG}). In any case, the single-exponential fit does capture
the relevant value for $T_2^{-1}$ at all temperatures.

A point of great interest is the measurement of the TSR at $T=20$ mK
in a FC and a ZFC sample, as shown in the inset of Fig.
\ref{T2FCZFC}. The decay of the transverse magnetization is best
fitted by Eq. (\ref{T2LG}), whereby the Gaussian component,
$T_{2G}^{-1}$, is separated from the Lorentzian one, $T_{2L}^{-1}$.
From the Gaussian component of the decay we can extract directly the
effect of the nuclear dipole-dipole interaction, whereas the other
mechanisms of dephasing (e.g. random changes in the local field due
to tunneling molecules) contribute mainly to the Lorentzian part.
The fit yields $T_{2G}^{-1}(\mathrm{FC}) = 104 \pm 3$ s$^{-1}$ and
$T_{2G}^{-1}(\mathrm{ZFC}) = 77 \pm 3$ s$^{-1}$. These results can
be understood by assuming that, at very low $T$, the main source of
TSR is the dipole-dipole coupling of like nuclei in neighboring
molecules. Then we can estimate $T_2^{-1}$ from the Van Vleck
formula for the second moment $M_2 = \langle \Delta \omega^2
\rangle$ of the absorption line in dipolarly-coupled
spins:\cite{vanvleck48PR}
\begin{eqnarray}
M_2 = \left(\frac{\mu_0}{4 \pi}\right)^2 \frac{3}{4} \gamma_N^4 \hbar^2 I(I+1) \sum_{i>j} \frac{(1-3 \cos^2
\theta_{ij})^2}{r_{ij}^6}, \label{vanvleck}\\ \nonumber
T_2^{-1} = \sqrt{M_2},
\end{eqnarray}
yielding $T_2^{-1} = 131$ s$^{-1}$ if we take for $r_{ij}$ the
distance between centers of neighboring molecules. The estimated
$T_2^{-1}$ would obviously be much larger if one would consider the
coupling between nuclei within the same cluster. As we argued when
discussing the $^{55}$Mn spectra, it is possible that the cluster
symmetry is low enough to prevent intracluster nuclear spin
flip-flops. This may explain why Eq. (\ref{vanvleck}) yields the
right order of magnitude when only coupling between nuclei in
neighboring molecules is considered. An alternative argument is
that, given the small number (4 at best) of like $^{55}$Mn spins
within one cluster, the dipolar coupling between them does not yield
a genuine decay of the transverse magnetization for the entire
sample. The macroscopic $T_2$ decay measured in the experiment
reflects therefore the slower, but global, intercluster spin
diffusion rate. A similar observation was recently made also in a
different molecular compound, Al$_{50}$C$_{120}$H$_{180}$ (Ref.
\onlinecite{bono07JCS}).

We also note that, in the case of a ZFC sample, the sum in Eq.
(\ref{vanvleck}) should be restricted to only half of the
neighboring molecules, since on average half of the spins have
resonance frequency $+\omega_{\rm N}$ and the other half
$-\omega_{\rm N}$, and no flip-flops can occur between nuclei
experiencing opposite hyperfine fields. This is equivalent to
diluting the sample by a factor 2, which reduces the expected
$T_2^{-1}$ in ZFC sample by a factor $\sqrt{2}$. Indeed, we find in
the experiment $T_{2G}^{-1}(\mathrm{FC})/T_{2G}^{-1}(\mathrm{ZFC}) =
1.35 \simeq \sqrt{2}$ which, together with the good quantitative
agreement with the prediction of Eq. (\ref{vanvleck}), constitutes
solid evidence for the presence of intercluster nuclear spin
diffusion. This is precisely the mechanism required to explain why
the tunneling in a minority of FRMs can relax the whole nuclear spin
system. The need for intercluster nuclear spin diffusion could
already have been postulated by analyzing the LSR rate, and the
magnetization dependence of the TSR rate gives an independent
confirmation.

For comparison, in a recent study of the $^{57}$Fe NMR in Fe$_8$, Baek \textit{et al.}\cite{baek05PRB} attributed
the observed TSR rate to the dipolar interaction between $^{57}$Fe and $^1$H nuclei. They analyzed their data
with the expression $T_2^{-1} \simeq (M_2^{(\rm H)} / 12 \tau_c)^{1/3}$, where $\tau_c$ is the proton TSR time
due to their mutual dipolar coupling and $M_2^{(\rm H)}$ is the second moment of the $^{57}$Fe - $^1$H coupling.
However, the same model\cite{takigawa86JPSJ} predicts the echo intensity to decay as $M_{\perp}(t)/M_{\perp}(0)
\simeq \exp(-2M_2^{(\rm H)} t^3 / 3)$. This function fails completely in fitting our echo decays, therefore we do
not consider the $^{55}$Mn - $^1$H dipolar coupling as an alternative explanation for the TSR we observe.

Finally we stress that, in our view, the fact that the LSR and the TSR are both roughly $T$-independent below 0.8
K does not find its origin in the same mechanism. Rather, we attribute them to two different mechanisms, both
$T$-independent: the quantum tunneling of the electron spin (for the LSR) and the nuclear spin diffusion (for the
TSR).

Having argued that the LSR in Mn$_{12}$-ac in driven by tunneling
fluctuations of the FRMs, which are peculiar of the acetate
compound, it's interesting to note that other varieties of Mn$_{12}$
molecules have meanwhile become available. In particular the
Mn$_{12}$-\textit{t}BuAc\cite{soler03CC,wernsdorfer06PRL} is a truly
axially symmetric variety that does not contain any FRMs, and could
provide an interesting counterexample for our results if studied by
low-$T$ NMR. The Mn$_{12}$BrAc molecule is also thought to be free
of FRMs,\cite{harter05IC} and some low-$T$ NMR experiments have been
performed on it\cite{chakov06JACS} that show indeed very different
results from what we report here. However, as we shall argue in
\S\ref{spinT}, a definite conclusion on the meaning of NMR
experiments at very low $T$ should only be drawn when the analysis
of the nuclear spin thermalization is included.

\subsection{Field dependence of the longitudinal spin relaxation rate} \label{fielddep}

Further insight in the interplay between the quantum tunneling fluctuations and the nuclear spin dynamics is
provided by the study of the dependence of the LSR on a magnetic field $B_z$ applied along the anisotropy axis.
It is clear from the Hamiltonian (\ref{hamiltonianMn12}) that, in the absence of other perturbations, such a
longitudinal field destroys the resonance condition for electron spin states on opposite sides of the barrier and
therefore inhibits the quantum tunneling. In the presence of static dipolar fields, $B_{\mathrm{dip}}$, by
studying the tunneling rate as a function of $B_z$ one may in principle obtain information about the distribution
of longitudinal $B_{\mathrm{dip}}$, since at a given value of $B_z$ there will be a fraction of molecules for
which $B_{\mathrm{dip}} = -B_z$ and will therefore be allowed to tunnel just by the application of the external
bias.

\begin{figure}[t]
\includegraphics[width=8cm]{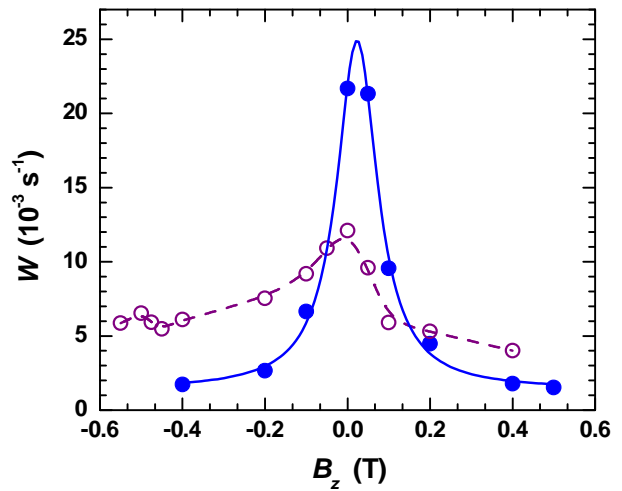}
\caption{\label{WvsB-lt} (Color online) Longitudinal field dependence of the LSR rate $W$ in the ZFC (solid dots)
and FC (open dots) sample at $T=20$ mK. The measuring frequencies are $\nu(B_z) = 230 + \gamma_N B_z$ MHz. The
solid line is a Lorentzian fit with HWHM $\Delta B_z \simeq 60$ mT. The dotted line through the FC data is a
guide for the eye.}
\end{figure}

We show in Fig. \ref{WvsB-lt} the LSR rate $W(B_z)$ at $T=20$ mK in
the ZFC sample, obtained while shifting the measurement frequency as
$\nu (B_z) = \nu (0) + \gamma_N B_z$ with $\nu(0)=230$ MHz, in order
to stay on the center of the NMR line that corresponds to the
molecules that are aligned exactly parallel with the applied field.
Since for a ZFC sample the magnetization is zero, the field
dependence should be the same when $B_z$ is applied in opposite
directions, as is observed. The data can be fitted by a Lorentzian
with a half width at half maximum (HWHM) $\Delta B_z \simeq 60$ mT:
this differs both in shape (Gaussian) and in width ($\Delta B_z
\simeq 21$ mT) from the calculated dipolar bias distribution in a
ZFC sample.\cite{tupitsynP} An alternative experimental estimate,
$\Delta B_z \simeq 25$ mT, can be found in magnetization relaxation
experiments,\cite{wernsdorfer99EPL} but only around the first level
crossing for FRMs ($\simeq 0.39$ T) in the FC sample. For
comparison, Fig. \ref{WvsB-lt} also shows $W(B_z)$ in the FC sample:
the shape is now distinctly asymmetric, with faster relaxation when
the external field is opposed to the sample magnetization.
Interestingly, $W(B_z)$ in the FC sample falls off much more slowly
on the tails for both positive and negative fields, while the value
at zero field is less than half that for the ZFC sample. We
therefore observe that in zero field the recovery of longitudinal
magnetization in the FC sample is faster than in the ZFC, whereas
the opposite is true for the decay of transverse magnetization
(inset Fig. \ref{T2FCZFC}).

\begin{figure}[t]
\includegraphics[width=8cm]{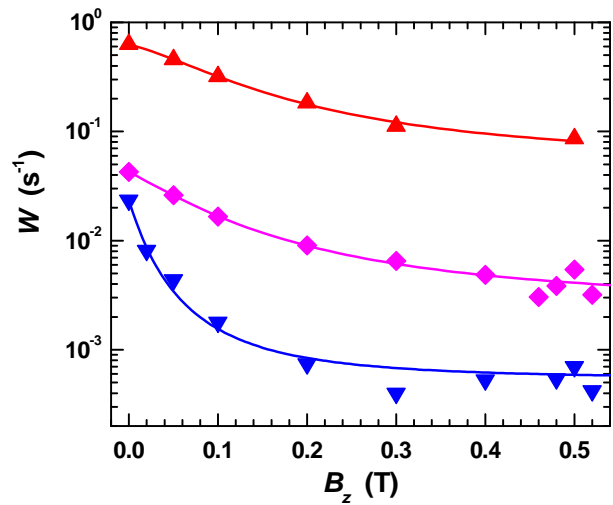}
\caption{\label{WvsB-T} (Color online) Longitudinal field dependence of the LSR rate in ZFC sample at $T=20$ mK
(down triangles), $T=720$ mK (diamonds) and $T=1.13$ K (up triangles). The measuring frequency in these datasets
is $\nu = 231 + \gamma_N B_z$ MHz. The lines are Lorentzian fits yielding HWHM $\Delta B_z = 16,85,118$ mT,
respectively.}
\end{figure}

If the LSR rate $W(B_z)$ is to be interpreted as a signature of quantum tunneling, its HWHM is clearly larger
than expected. Part of the reason may be the fact that the width of the Mn$^{(1)}$ line is already intrinsically
larger than both $\Delta B_z$ and the distribution of dipolar fields created by the molecules. Indeed, the width
of the Mn$^{(1)}$ line, $\sigma_{\nu} \simeq 1.2$ MHz, translates into a local field distribution of width
$\sigma_{\rm B} \simeq 115$ mT for $^{55}$Mn. The observed HWHM does depend, for instance, on the choice of
$\nu(0)$. As soon as $B_z \neq 0$ the presence of slightly misaligned crystallites in our sample may also
contribute to the width of the resonance. In any case, all of the mechanisms mentioned above (distribution of
internal dipolar fields, width of the NMR line, distribution of crystallite orientations in the sample) would
yield a $T$-independent linewidth for $W(B_z)$. Fig. \ref{WvsB-T} shows $W(B_z)$ in ZFC sample at three different
temperatures, $T=0.02,0.72,1.13$ K, covering the pure quantum regime, the thermally-activated regime, and the
crossover temperature. The NMR frequency in these datasets is $\nu (B_z) = 231 + \gamma_N B_z$. The data have
been fitted by Lorentzian lines yielding a HWHM $\Delta B_z = 16,85,118$ increasing with temperature. We note
immediately that the HWHM at $T=20$ mK is much smaller than the one obtained from the data in Fig. \ref{WvsB-lt},
the only difference between the two sets being $\nu(0)$ and, subsequently, all other measurement frequencies at
$B_z \neq 0$. Indeed, we found that already in zero field the LSR rate does depend on $\nu$, reaching the highest
values at the center of the line and falling off (up to a factor 5) on the sides. This dependence, however,
becomes much weaker at high temperatures. It is therefore rather difficult to make strong statements about the
meaning of the observed increase in $\Delta B_z$ with temperature. At any rate, however, the field dependencies
observed here at low-$T$ are much stronger than those previously reported in the high-$T$
regime.\cite{furukawa01PRB,goto03PRB} Goto \textit{et al.} also reported $W(B_z)$ for the ``lower branch'' of the
Mn$^{(1)}$ line, viz. for the nuclei whose local hyperfine field is opposite to the external field (Ref.
\onlinecite{goto03PRB}, Fig. 6, closed squares). That situation is equivalent to our FC data (Fig. \ref{WvsB-lt},
open dots) for $B_z < 0$. At large fields an overall increase of $W$ with $B_z$ is observed in Ref.
\onlinecite{goto03PRB}, but for $B_z < 1$ T the LSR rate does decrease, in agreement with our results.

We also noted, both in Fig. \ref{WvsB-T} and in the FC data in Fig.
\ref{WvsB-lt}, that a small increase in $W(B_z)$ occurs at $|B_z|
\simeq 0.5$ T, which is approximately the field value at which the
$|+9\rangle$ and $|-10\rangle$ electron spin states come into
resonance. This feature is barely observable, but nevertheless well
reproducible. As a counterexample, in another dataset (not shown) we
investigated $W(B_z)$ more carefully in the FC sample at $T=20$ mK
for positive values of $B_z$, and found no increase around $B_z
\simeq 0.5$ T, as one would expect since the fully populated state,
$|-10\rangle$, is pushed far from all other energy levels. A
similarly small peak in $W(B_z)$ at the first levels crossing has
been recently observed in Fe$_8$ as well.\cite{baek05PRB}

\subsection{Deuterated sample} \label{deuterated}

The role of the fluctuating hyperfine bias on the incoherent tunneling dynamics of SMMs, predicted by Prokof'ev
and Stamp,\cite{prokof'ev96JLTP} has been clearly demonstrated by measuring the quantum relaxation of the
magnetization in Fe$_8$ crystals in which the hyperfine couplings had been artificially modified by substituting
$^{56}$Fe by $^{57}$Fe or $^1$H by $^2$H (Ref. \onlinecite{wernsdorfer00PRL}). For instance, the time necessary
to relax 1\% of the saturation magnetization below 0.2 K was found to increase from 800 s to 4000 s by
substituting protons by deuterium, whereas it decreased to 300 s in the $^{57}$Fe enriched sample. More recently,
Evangelisti \emph{et al.}\cite{evangelisti05PRL} showed that the $^{57}$Fe isotopic enrichment of Fe$_8$ causes
the magnetic specific heat to approach its equilibrium value within accessible timescales ($\sim 100$ s).

\begin{figure*}[t]
\includegraphics[width=11cm]{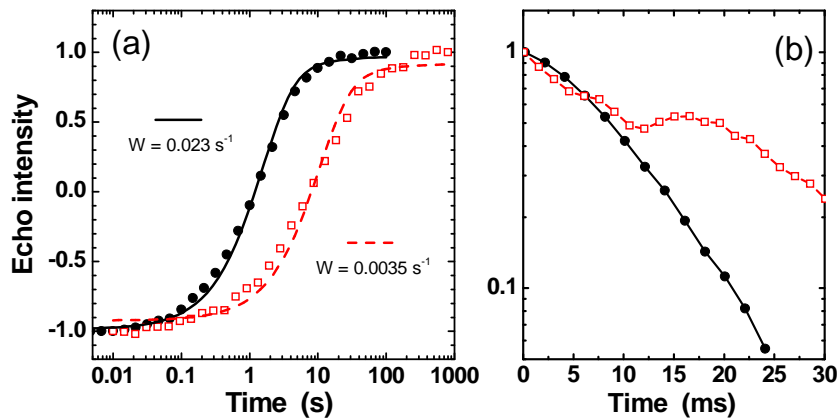}
\caption{\label{T1T2D} (Color online) Comparison between (a) the
nuclear inversion recoveries and (b) the decays of transverse
magnetization in the "natural" Mn$_{12}$-ac (circles) and in the
deuterated sample (squares), at $T=20$ mK in zero field and ZFC
sample, for the Mn$^{(1)}$ site. The solid lines in (a) are fits to
Eq. (\ref{recovery}).}
\end{figure*}

Since in Mn$_{12}$-ac the only possible isotope substitution is
$^1$H $\rightarrow ^2$H, we performed a short set of measurements on
a deuterated sample. The sample consists of much smaller
crystallites than the ``natural'' ones used in all other experiments
reported here. Although a field-alignment was attempted following
the same procedure as described in \S\ref{measures}, the orientation
of the deuterated sample turned out to have remained almost
completely random, probably due to the too small shape anisotropy of
the crystallites. We therefore report only experiments in zero
external field, where the orientation is in principle irrelevant.

The results are shown in Fig. \ref{T1T2D}: the $^{55}$Mn LSR rate at $T=20$ mK in zero field and ZFC sample is
indeed reduced to $W_{\mathrm{deut}} \simeq 0.0035$ s$^{-1}$, i.e. 6.5 times lower than in the ``natural''
sample. This factor is the same as the reduction of the electron spin relaxation rate seen in deuterated Fe$_8$
(Ref. \onlinecite{wernsdorfer00PRL}), and it coincides with the ratio of the gyromagnetic ratios of $^1$H and
$^2$H. This finding unequivocally proves that the proton spins are very effective in provoking the tunneling
events via the Prokof'ev-Stamp mechanism, and confirms that the LSR rate of the $^{55}$Mn nuclei is a direct
probe of the electron spin tunneling rate.

As regards the TSR, the result is quite intriguing: slow but rather ample oscillations are superimposed to the
decay of transverse magnetization, and the overall decay rate appears slower than in the natural sample. This
behavior is reminiscent of the change in TSR rate upon application of a small longitudinal magnetic field in the
natural sample. The latter has a rather complicated physical origin and is still under investigation.

\subsection{Comparison with a Mn$^{3+}$ site}  \label{mn3+}

\begin{figure*}[t]
\includegraphics[width=11cm]{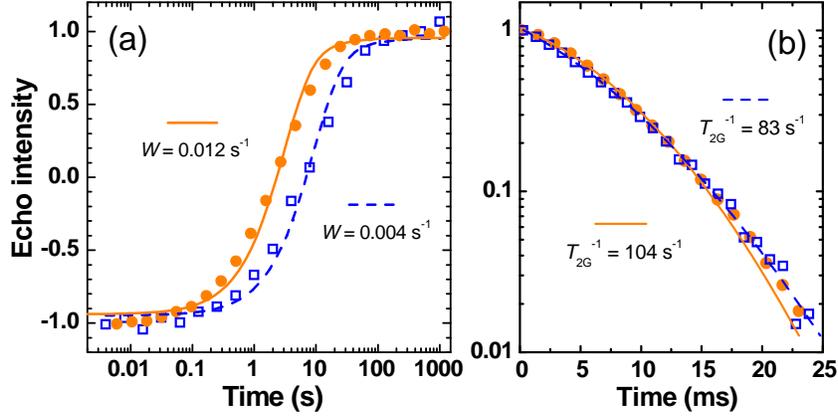}
\caption{\label{T1T2Mn34}  (Color online) Comparison between (a) the
recovery of longitudinal magnetization and (b) the decay of
transverse magnetization in Mn$^{(1)}$ (circles) and Mn$^{(2)}$
(diamonds) sites, at $T=20$ mK in FC sample and zero external field.
The solid (Mn$^{(2)}$) and dashed (Mn$^{(1)}$) lines are fits to
Eq.(\ref{recovery}) in panel (a) and Eq.(\ref{T2LG}) in panel (b).}
\end{figure*}

\begin{figure}[b]
\includegraphics[width=8cm]{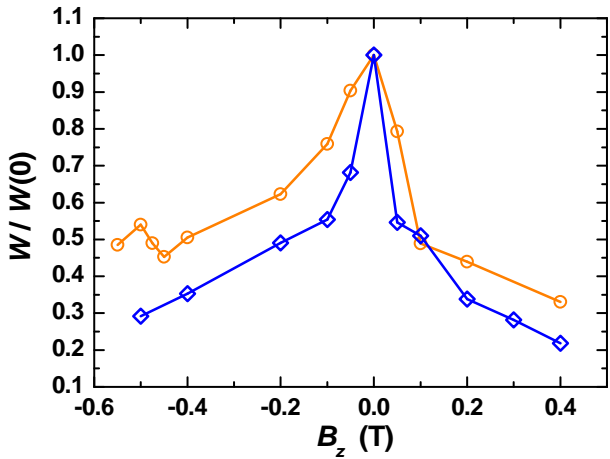}
\caption{\label{WvsB34}  (Color online) Longitudinal field dependencies of the LSR rates in Mn$^{(1)}$ (circles)
and Mn$^{(2)}$ (diamonds) sites, normalized at the zero-field value. The data are taken at $T=20$ mK in FC sample
with central measuring frequencies $\nu^{(1)}(0) = 230$ MHz and $\nu^{(2)}(0) = 283.7$ MHz.}
\end{figure}

Some rather interesting results emerge from the analysis of extra measurements performed on the NMR line of the
Mn$^{(2)}$ site, i.e. a Mn$^{3+}$ ion. Fig. \ref{T1T2Mn34} shows a comparison between the recovery of the
longitudinal magnetization and the decay of the transverse magnetization in Mn$^{(1)}$ and Mn$^{(2)}$ sites, at
$T=20$ mK in the FC sample and zero external field, at a frequency $\nu^{(2)} = 283.7$ MHz. The TSR is very
similar in both sites, although a closer inspection evidences that the Gaussian nature of the decay is less
pronounced in the Mn$^{(2)}$ sites, which leads to $T_{2G}^{-1} = 83$ s$^{-1}$ instead of the $T_{2G}^{-1} = 104$
s$^{-1}$ found in Mn$^{(1)}$. More importantly, the LSR is three times slower in the Mn$^{(2)}$ site, as seen in
Fig. \ref{T1T2Mn34}(b). This is opposite to the high-$T$ regime, where the Mn$^{3+}$ sites were
found\cite{furukawa01PRB,goto03PRB} to have much faster relaxation. Furthermore, the field dependence of the LSR
rate appears sharper in the Mn$^{(2)}$ site, as shown in Fig. \ref{WvsB34}. The asymmetry in $W(B_z)$ for a FC
sample is still present, but less evident than in the Mn$^{(1)}$ site due to the more pronounced decrease of $W$
already for small applied fields.

The similarity between the TSR rates in the Mn$^{(1)}$ and the Mn$^{(2)}$ sites is indeed expected if $T_2$ is
determined by intercluster nuclear spin diffusion. Conversely, the difference in LSR is more difficult to
understand if one assumes that the process that induces longitudinal spin relaxation is the tunneling of the
molecular spin. However, one clear difference between Mn$^{(1)}$ and Mn$^{(2)}$ is the width of the NMR line,
much larger in Mn$^{(2)}$. Since the integrated intensity of both lines is identical, the Mn$^{(2)}$ has an
accordingly lower maximum intensity. We have verified for both sites that the LSR rate is the fastest when
measuring at the highest intensity along each line. Thus, the factor 3 slower LSR in Mn$^{(2)}$ could simply be
another manifestation of the apparent dependence of the measured $W$ on the NMR intensity along each line. We
point out, however, that the measured LSR rate is independent of the $\pi/2$ pulse length, which determines the
spectral width of the pulse and thereby the fraction of spins being manipulated and observed. This means that the
difference in W for the two sites cannot be simply attributed to a difference in the number of spins excited
during a pulse of given length but that other (more complex) factors must play a role.

\section{Thermalization of the nuclear spins} \label{spinT}

Having demonstrated that the $^{55}$Mn longitudinal spin relaxation
below 0.8 K is driven by $T$-independent quantum tunneling
fluctuations, a natural question to ask is whether or not the
nuclear spins are in thermal contact with the lattice at these low
temperatures. Let us recall that any direct coupling between phonons
and nuclear spins is expected to be exceedingly weak, due to the
very small density of phonons at the nuclear Larmor
frequency.\cite{abragam61} Relaxation through electric quadrupole
effects, if present, would show a temperature dependence $\propto
(T/\Theta_{\rm D})$ for direct process or $\propto (T/\Theta_{\rm
D})^2$ for Raman process ($\Theta_{\rm D}$ is the Debye
temperature), which is not consistent with our observations.
Therefore the thermalization of the nuclei will have to take place
via the electron spin - lattice channel. Since in the quantum regime
the only electron spin fluctuations are due to tunneling, the
question whether the nuclear spins will still be in equilibrium with
the lattice temperature is of the utmost importance.

\begin{figure}[t]
\includegraphics[width=8.5cm]{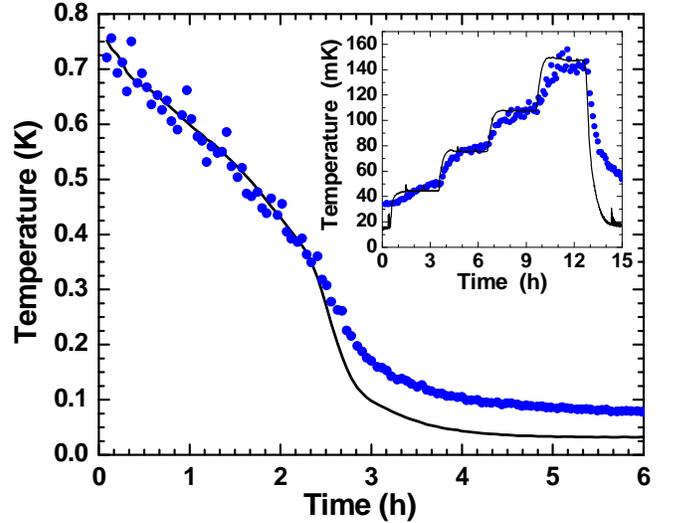}
\caption{\label{Tnuclcooldown} (Color online) Comparison between
bath temperature $T_{\mathrm{bath}}$(solid lines) and nuclear spin
temperature $T_{\mathrm{nucl}}$ (circles), while cooling down the
system (main panel) and while applying step-like heat loads (inset).
The waiting time between NMR pulses was 60 s in the main panel and
180 s in the inset. Both datasets are at zero field in ZFC sample.}
\end{figure}

\subsection{Time evolution of the nuclear spin temperature}
\label{timespinT}

We have addressed this problem by cooling down the refrigerator from
800 to 20 mK while monitoring simultaneously the temperature
$T_{\mathrm{bath}}$ of the $^3$He/$^4$He bath in the mixing chamber
(just next to the sample) and the NMR signal intensity of the
Mn$^{(1)}$ line, in zero external field and on a ZFC sample. The
signal intensity was measured by spin echo with repetition time
$t_{\rm rep} = 60$ s. The nuclear spin temperature\cite{goldman70}
$T_{\mathrm{nucl}}$ is obtained as described in \S\ref{measures},
and plotted in Fig. \ref{Tnuclcooldown} together with
$T_{\mathrm{bath}}$. We find that the nuclear spin temperature
strictly follows the bath temperature, with small deviations
starting only below $\sim 200$ mK. This result is crucial but rather
paradoxical, and we shall discuss its implications in detail in
\S\ref{thermaleq}. Experimentally, however, it certifies the
effectiveness of our cryogenic design in achieving the best possible
thermalization of the sample, since the nuclear spins are the last
link in the chain going from the $^3$He/$^4$He bath via the phonons
in the sample to the electron spins and finally to the nuclei.

The lowest spin temperature that can be \textit{measured} appears to
depend on the pulse repetition time $t_{\rm rep}$. To measure
$T_{\mathrm{nucl}}$ with the pulse NMR method we need a $\pi/2$
pulse to create a transverse nuclear magnetization, and after a time
$T_2$ the spins are effectively at infinite $T$ so enough time must
elapse before taking the next $T_{\mathrm{nucl}}$ measurement. For
the data in Fig. \ref{Tnuclcooldown}, $t_{\rm rep} = 60$ s was
barely longer than the observed time for inversion recovery [see
Fig. \ref{echo3D}(b)], and the lowest observed spin temperature is
$T_{\mathrm{nucl}}^{\mathrm{min}} \simeq 80$ mK. This improved when
using longer waiting times between pulses, e.g.
$T_{\mathrm{nucl}}^{\mathrm{min}} \simeq 35$ mK with $t_{\rm rep} =
180$ s, as shown in the inset of Fig. \ref{Tnuclcooldown}. However,
no matter how long the waiting time, we never observed a
$T_{\mathrm{nucl}}$ lower than $\sim 30$ mK.

\begin{table}
\caption{\label{tableTnucl}Experimental conditions and relaxation
rates for the nuclear spin temperature experiments in Fig.
\ref{spinTcompare} }
\begin{ruledtabular}
\begin{tabular}{cccccccc}
Panel & Mn  & $B_z$ & $\dot{n}$ & $\dot{Q}$ & $t_{\rm rep}$ & $T_1$ & $\tau_{\rm th}$\\
 & site & (T) & ($\mu$mol/s) & (mW) & (s) & (s) & (min)\\
\hline
a & 1 & 0 &  330 & 0.63 & 120 & 41.3 & $58 \pm 5$\\
b & 2 & 0 &  330 & 0.63 & 120 & 122 & $83 \pm 13$ \\
c & 1 & 0 &  430 & 0.78 & 120 & 41.3 & $37 \pm 3$\\
d & 2 & 0.2 & 330 & 0.63 & 300 & 355 & $92 \pm 33$
\end{tabular}
\end{ruledtabular}
\end{table}

\begin{figure*}[t]
\includegraphics[width=10cm]{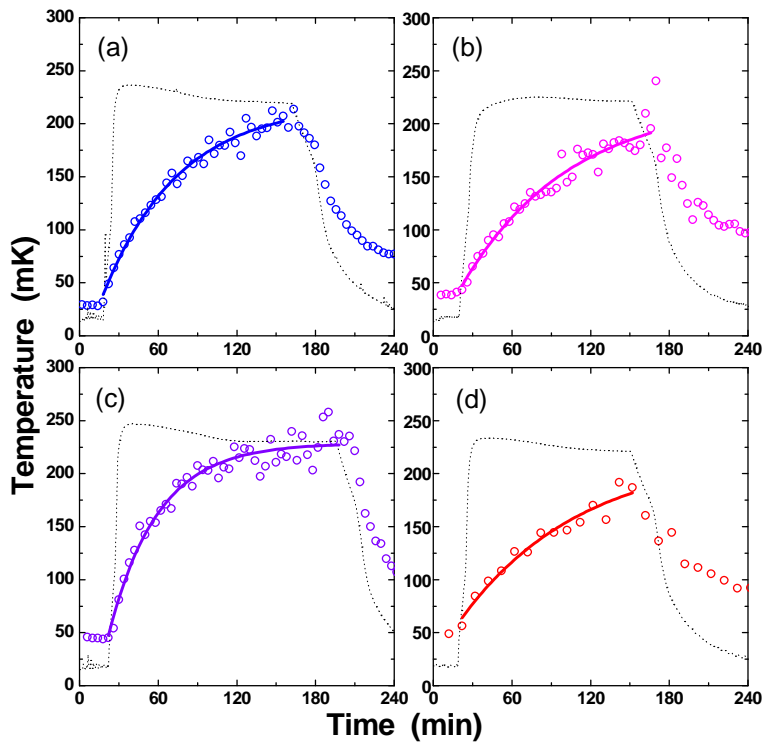}
\caption{\label{spinTcompare}(Color online) Time evolution of the
nuclear spin temperature (open symbols) and the bath temperature
(dotted lines) upon application of a step-like heat load. All data
are for a FC sample. The solid lines are fits to Eq. (8), yielding
the thermal time constants $\tau_{\mathrm{th}}$ reported in table
\ref{tableTnucl}, along with the Mn site, external magnetic field
$B_z$, LSR rate $W$, NMR pulse repetition time $t_{\rm rep}$,
$^{3}$He flow rate $\dot{n}$, and applied heat load $\dot{Q}$.
Notice in particular the effect of a change in $^3$He circulation
rate, panel (c) vs. panel (a).}
\end{figure*}

Next we study the time constant $\tau_{\mathrm{th}}$ for the thermalization of the nuclear spin system with the
helium bath, by applying step-like heat loads and following the time evolution of $T_{\mathrm{nucl}}$. In
particular, we are interested in the relationship between $\tau_{\mathrm{th}}$, the LSR time $T_1 = 1/2W$ as
obtained from the inversion recovery technique, and the $^3$He circulation rate $\dot{n}$, which is proportional
to the refrigerator's cooling power, $\dot{Q}$. $T_1$ is easily tuned by measuring at different longitudinal
fields and Mn sites, while $\dot{n}$ is changed by applying extra heat to the refrigerator still. Since also the
NMR signal intensity changes under different fields and Mn sites, we must redefine every time the conversion
factor $K$ between signal intensity and $T_{\mathrm{nucl}}$. In the following we choose $K$ such that the
asymptotic value of $T_{\mathrm{nucl}}$ for $t \rightarrow \infty$ matches the measured $T_{\mathrm{bath}}$ at
the end of the heat step. This implies the assumption that the measuring pulses do not saturate, i.e. ``heat
up'', the nuclear spins, and requires $T_{\rm rep} > T_1$. Fig. \ref{spinTcompare} shows four examples of the
time evolution of $T_{\mathrm{nucl}}$ under the application of a heat load for $\sim 2$ hours, in Mn$^{(1)}$ and
Mn$^{(2)}$ sites, with or without an applied field, and with an increased $^{3}$He flow rate. We fitted the data
to the phenomenological function:
\begin{eqnarray}
T_{\mathrm{nucl}}(t) = T_{\mathrm{nucl}}(0) +  \\ \nonumber
 [T_{\mathrm{nucl}}(\infty) - T_{\mathrm{nucl}}(0)]
\left[1 - \exp \left(-\frac{t - t_0}{\tau_{\mathrm{th}}}\right)\right], \label{Tspinexp}
\end{eqnarray}
where $T_{\mathrm{nucl}}(\infty)$ is set by definition equal to
$T_{\mathrm{bath}}$ at the end of the step, $T_{\mathrm{nucl}}(0)$
follows automatically from the above constraint, and $t_0$ is the
time at which the heat pulse is started. We find that
$\tau_{\mathrm{th}}$ is always much longer than the nuclear LSR time
$T_1$, and that larger $T_1$ corresponds to larger
$\tau_{\mathrm{th}}$. However, the dependence of
$\tau_{\mathrm{th}}$ on Mn site and applied field is not as strong
as for $T_1$, i.e. $\tau_{\mathrm{th}}$ and $T_1$ are not strictly
proportional to each other. Conversely, by changing the $^{3}$He
flow rate we observe that, within the errors, the ratio of heat
transfer from the $^{3}$He stream to the nuclear spins is
proportional to $\dot{n}$, given the same conditions of nuclear site
and external field.

We should stress that, when measuring $T_1$ by inversion recovery, we effectively ``heat up'' only a small
fraction of the nuclear spins, namely those whose resonance frequencies are within a range, $\delta \nu$,
proportional to the inverse of the duration, $t_{\pi}$, of the $\pi$-pulse. With $t_{\pi} \simeq 20$ $\mu$s we
get $\delta \nu = 1/2 \pi t_{\pi} \simeq 8$ KHz, which is less than 0.2\% of the width of the Mn$^{(1)}$ line.
Conversely, by increasing the bath temperature we heat up the entire spin system, thereby requiring a much larger
heat flow to occur between the $^3$He stream and the nuclear spins. Therefore, these results show that the
thermal equilibrium between nuclear spins and lattice phonons does occur on a timescale of the order of $T_1$ as
obtained from inversion recovery, since the main bottleneck appears to be between lattice phonons and $^{3}$He
stream, as demonstrated by the dependence of $\tau_{\mathrm{th}}$ on $\dot{n}$. In a later set of experiments
(not shown here) using a small single crystal instead of a large amount of oriented powder, we have indeed
observed an even shorter $\tau_{\mathrm{th}}$, which indicates that $\tau_{\mathrm{th}}$ should ultimately tend
to $T_1$ for small sample size and strong thermal contact between lattice phonons and helium bath.

\subsection{Longitudinal field sweeps and magnetic avalanches}
\label{superradiance}

To conclude our study on the nuclear spin thermalization, we
attempted to measure $T_{\mathrm{nucl}}$ in the presence of large
longitudinal magnetic field sweeps, motivated by the fact that much
of the experiments on spin tunneling in SMMs are based on the
measurement of magnetic hysteresis loops. Under those conditions,
the electron spins are flipped at abnormally large rates, and one
may ask whether or not the nuclear spins are still able to remain in
thermal equilibrium. Unfortunately, monitoring $T_{\mathrm{nucl}}$
while $B_z$ is being swept means that one should continuously change
the NMR probe frequency, and synchronize that change with the field
sweep. This being technically cumbersome, we could only measure
$T_{\mathrm{nucl}}$ at zero field at the beginning and at the end of
a $B_z$ sweep. The results are somewhat inconclusive and shall not
be discussed here, but more details can be found in section 4.4.2 of
Ref. \onlinecite{morelloT}.

\begin{figure}[t]
\includegraphics[width=8.5cm]{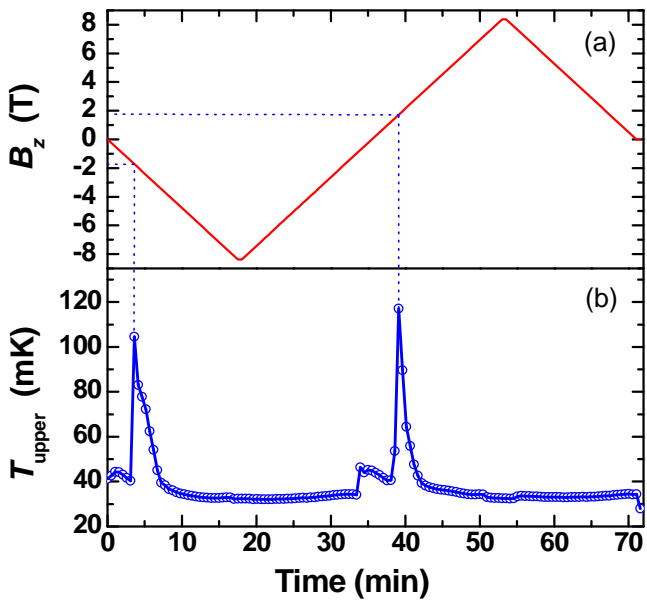}
\caption{\label{avalanches}(Color online) (a) Longitudinal magnetic
field and (b) temperature of the upper thermometer (see Fig.
\ref{dilution+temperatures}) during a field sweep at $\mathrm{d}B_z
/ \mathrm{d}t = 0.5$ T/min. The sample was initially field-cooled
with $B_z > 0$. The sharp jumps in $T_{\rm upper}$ occur when $|B_z|
\simeq 1.9$ T, i.e. at the fourth level crossing field, and are
attributed to the radiation produced by a magnetic avalanche.}
\end{figure}

We do mention, however, that during the $B_z$-sweep experiments we
always encountered magnetic avalanches, i.e., abrupt reversal of the
electronic magnetization of the whole sample. This phenomenon has
been first reported already some time ago\cite{paulsen95JMMM} but is
only recently being studied in more detail.\cite{suzuki05PRL}
Importantly, the magnetization reversal is expected to be
accompanied by the emission of electromagnetic
radiation,\cite{tejada04APL} which is in fact what we observed in
our experiments, since we were not equipped to measure the
electronic magnetization directly on short time scales. Fig.
\ref{avalanches} shows the temperature recorded by the upper
thermometer in the mixing chamber (see Fig.
\ref{dilution+temperatures}), while the longitudinal field is being
swept at a rate $\mathrm{d}B_z / \mathrm{d}t = 0.5$ T/min. The
sweeping field gives a heat load that raises the observed
temperature to $\simeq 30$ mK, but the most striking feature of the
data is the sudden jump of $T_{\rm upper}$ to above 100 mK, whenever
the applied field reaches $|B_z| \simeq 1.9$ T \emph{and} its
direction is opposite to the instantaneous magnetization. We note
that the timescale for the apparent temperature jump is essentially
identical to what we observe immediately after the application of a
rf-pulse for NMR measurements, as shown in Fig.
\ref{dilution+temperatures}(b). In the same figure it is seen that a
heat pulse applied at the sample location shows its effect at the
upper thermometer with a delay of about 3 minutes (due to the $^3$He
drift velocity) in the form of a broad temperature ``bump''. We
therefore conclude that the sudden jumps in $T_{\rm upper}$ shown in
Fig. \ref{avalanches}(b) must be of \emph{electromagnetic} rather
than \emph{thermal} origin, and may be attributed to the radiation
produced by the sudden reversal of the entire electronic
magnetization of the sample by the magnetic
avalanche.\cite{tejada04APL} The radiation bursts reported in Ref.
\onlinecite{tejada04APL} at a temperature $T=1.8$ K occurred at
$|B_z^{\rm (av)}| \simeq 1.4$ T, which corresponds to the third
level crossing field for spin tunneling, i.e., the value of field at
which the resonance between $m = \pm 10$ and $m = \mp 7$ states is
obtained. We found instead the avalanches at $|B_z^{\rm (av)}|
\simeq 1.9$ T, i.e., the fourth level crossing, $m = \pm 10
\leftrightarrow \mp 6$, but our measurements are done at $T \simeq
30$ mK. Goto \textit{et al.}\cite{goto03PhyB} also reported the
observation of magnetic avalanches in Mn$_{12}$-ac, and studied the
temperature dependence of the avalanche field $B_z^{\rm (av)}$.
Their finding that $B_z^{\rm (av)}$ increases with temperature was
interpreted as a sign that the avalanches occur more easily when the
thermal contact to the bath is weaker. Indeed, whereas they would
observe avalanches even at fields as low as $B_z^{\rm (av)} \simeq
0.5$ T (the first level crossing) with the sample loosely anchored
to the mixing chamber of a dilution refrigerator at $T=0.15$ K, they
never saw avalanches when the same sample was placed directly in a
liquid helium bath at $T=1.4$ K. In this sense, our observation of a
high $B_z^{\rm (av)} \simeq 1.9$ T confirms once more that our
strategy for the sample thermalization is very effective. Suzuki
\textit{et al.}\cite{suzuki05PRL} found even higher values of
$B_z^{\rm (av)}$ at subkelvin temperatures when measuring the local
magnetization of a small Mn$_{12}$-ac crystal immersed in liquid
$^3$He. However, their observations differ markedly from ours in
that they found avalanches occurring in a wide range of (not
necessarily resonant) fields, whereas we saw avalanches always and
only at the fourth level-crossing field.

\section{Analysis of the nuclear spin dynamics and theoretical implications} \label{theory}

In this section we attempt a quantitative analysis of our
experimental results, particularly the observed values of LSR rate.
To this end, we shall apply the Prokof'ev-Stamp (PS) theory of the
spin bath, which describes the dynamics of a ``central spin''
$\mathbf{S}$ (here the giant electronic spin of a Mn$_{12}$-ac
cluster) coupled to a bath of environmental (in this case, nuclear)
spins. In view of the complexity of the model we provide here an
introductory overview of some essential elements of the PS theory
needed for our analysis, referring the reader to the original
papers\cite{prokof'ev95CM,prokof'ev96JLTP,prokof'ev00RPP,stamp04CP}
for more details. For comparison, we also calculate the LSR rate
assuming that the electron spin tunneling is driven by spin-phonon
coupling.\cite{kagan80JETP,stamp04CP} We anticipate that the result
of this effort will be that the existing theory is not sufficient to
properly describe these and other related
experiments.\cite{evangelisti04PRL,evangelisti05PRL} We shall carry
out the analysis in detail in order to emphasize at every step what
assumptions are being made, what is their actual validity, and why
the known theories cannot explaining the data.

The goal of our analysis is to link the electron spin tunneling rate, $\Gamma = \tau_{\rm T}^{-1}$, to the
observed LSR rate, $W$, based on the following assumptions, justified by the experiments presented in the
previous sections: (i) The nuclear relaxation is driven by tunneling fluctuations in a minority of fast-relaxing
molecules. We shall assume the fraction of FRMs to be 5\% of the total.\cite{wernsdorfer99EPL} The neighboring
slow molecules can be safely considered as frozen during the timescale of interest and serve simply as a
``reservoir of nuclear polarization''. (ii) The dipole-dipole coupling between $^{55}$Mn nuclei in equivalent
sites of neighboring molecules allows intercluster nuclear spin diffusion, at a rate $T_2^{-1}$ much faster than
the LSR rate. (iii) The nuclear spin system is in thermal equilibrium with the phonon bath.

Before we start, it is of interest to point out some rather striking
peculiarities of the problem at hand. First and most importantly,
one cannot use any result from perturbation theory here, because the
nuclear Zeeman splittings arise uniquely from hyperfine fields,
which themselves jump between two different directions each time the
electron spin of a molecule tunnels, so there is no static part of
the nuclear Hamiltonian. Perhaps the only situation that resembles
this is the nuclear quadrupolar relaxation in systems with molecular
rotations.\cite{alexander65PR} Conversely, in the overwhelming
majority of NMR experiments one has a static external field
(produced by an actual magnet) and some local fluctuating fields
arising from the magnetic environment of the nuclei, which can be
treated as small perturbations. Then the LSR rate is easily related
to the spectral density of the local magnetic fluctuations,
calculated at the NMR frequency determined by the external
field.\cite{abragam61,slichterB}  Also curious is the way nuclear
spin diffusion proceeds in our system. The well-known treatment of
nuclear relaxation by coupling to paramagnetic impurities plus
nuclear spin diffusion\cite{lowe68PR} shows that there is a ``spin
diffusion barrier radius'' below which neighboring nuclear spins
cannot exchange energy because the large dipolar field from the
impurity brings them out of resonance. Here, instead, there is no
such minimum radius for spin diffusion because nuclei at equivalent
sites of different molecules are also magnetically equivalent
(provided both molecules have the same electron spin orientation).

\subsection{Spin-bath analysis and tunneling rate} \label{spinbath}

\begin{figure}[t]
\includegraphics[width=6cm]{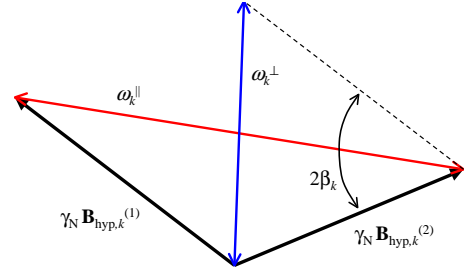}
\caption{\label{PSangles} (Color online) Scheme of the relative orientations of the hyperfine fields before
($\mathbf{B}_{{\rm hyp},k}^{(1)}$) and after ($\mathbf{B}_{{\rm hyp},k}^{(2)}$) the electron spin flip, and the
components of the hyperfine coupling that change ($\omega_k^{\parallel}$) or stay unchanged ($\omega_k^{\perp}$)
at each tunneling event. The angle $\beta_k$ is involved in the definition of $\kappa$, the number of nuclei
coflipping by ``orthogonality blocking'', Eq. (\ref{defkappa}).}
\end{figure}

\begin{figure*}[t]
\includegraphics[width=11cm]{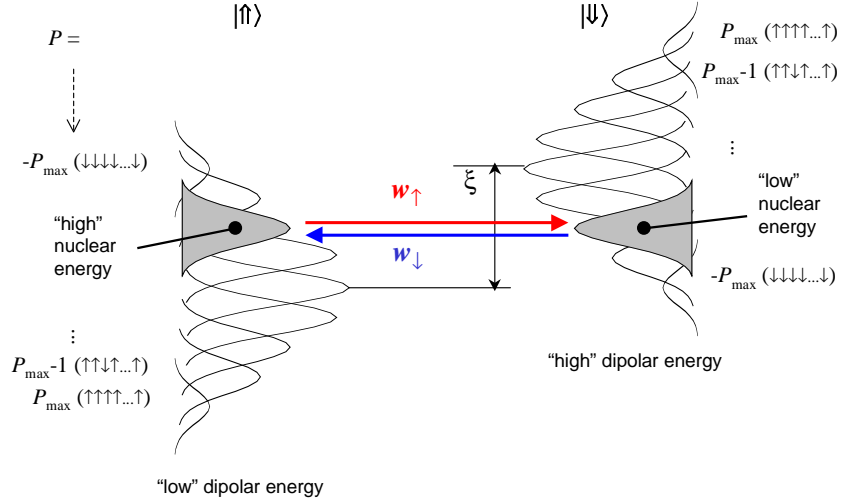}
\caption{\label{tunnelbias} (Color online) Sketch of the hyperfine-split manifolds representing the energy of the
$m = \pm S$ electron spin levels coupled to the nuclear spin bath.}
\end{figure*}

To apply the spin bath theory to the $^{55}$Mn NMR in Mn$_{12}$-ac,
we begin by truncating the giant spin Hamiltonian of the cluster to
its tunneling-split ground doublet, and by taking as a basis for its
subspace the $m =\pm S$ projections of $\mathbf{S}$ along the
$\hat{z}$-axis, denoted by $|\Uparrow\rangle,|\Downarrow\rangle$.
This restriction will be relaxed to consider higher excited electron
spin doublets when discussing thermally-assisted tunneling. Further,
we assume that each central spin is coupled to $N$ nuclear spins
$\{\mathbf{I}_k\}$, $k = 1 \ldots N$. The strength of each coupling
is given by the quantities $\hbar \omega_k^{\parallel}$ and $\hbar
\omega_k^{\perp}$, which represent the part of the hyperfine
coupling that does or does not change upon flipping the central
spin, respectively (Fig. \ref{PSangles}). For nuclei in Mn$^{(1)}$
sites of Mn$_{12}$-ac the hyperfine field $\mathbf{B}_{\rm hyp}$ is
exactly parallel or antiparallel to the direction of the cluster's
$\hat{z}$ axis, so $\omega_k^{\perp} = 0$ and $\omega_k^{\parallel}
= \gamma_N B_{\rm hyp}$. In Mn$^{(2)}$ and Mn$^{(3)}$ sites there's
a small nonzero value of $\omega_k^{\perp}$ due to the orbital
contribution to the hyperfine field.\cite{kubo02PRB} Conversely, for
nuclei such as $^1$H, which are subject to the vector sum of the
dipolar fields from several surrounding clusters, we may expect
$\omega_k^{\perp}$ and $\omega_k^{\parallel}$ to have comparable
values. Let us define for each nuclear spin a number $m_k$
representing the spin projection of $\mathbf{I}_k$ along the
direction of the local hyperfine field $\mathbf{B}_{{\rm hyp},k}$.
For $^1$H nuclei $m_k = \pm 1/2$, while for $^{55}$Mn $m_k = -5/2
\ldots +5/2$. Then the total hyperfine bias on the cluster is $\xi_N
= -2\hbar \sum_{k=1}^N m_k \omega_k^{\parallel}$. With this
definition, $\xi_N < 0$ when the majority of nuclear spins is
parallel to the local $\mathbf{B}_{{\rm hyp},k}$, thereby lowering
the total energy of the system. Notice that, for a given orientation
of the nuclear spins, $\xi_N$ changes sign whenever the electron
spin flips, since the direction of $\mathbf{B}_{\rm hyp}$ does. Thus
we define an \emph{absolute} index of nuclear polarization in each
cluster as $\mathcal{P} = C_S \sum_k m_k$, with $C_S = +1$ when
$\mathbf{S}$ is in the $|\Uparrow\rangle$ state, and $C_S = -1$
otherwise. Each possible value of $\mathcal{P}$ defines a
``polarization group'', and is independent of the electron spin
state. Since the individual hyperfine couplings vary over a broad
range (from $\sim 1$ MHz for distant protons to 365 MHz in
Mn$^{(3)}$), the possible values of the bias $\xi_N$ for each
$\mathcal{P}$ are also widely spread, yielding a set of largely
overlapping polarization groups. Globally, we may describe the
coupled ``central spin + spin bath'' system by two manifolds of
states, one for each electron spin state
$|\Uparrow\rangle,|\Downarrow\rangle$, split by hyperfine
interactions into a dense band of states indexed by the nuclear
polarization $\mathcal{P}$, as shown in Fig. \ref{tunnelbias}.
Calling $\mathcal{P}_{\rm max}$ the maximum value assumed by
$\mathcal{P}$, $\mathcal{P}_{\rm max} = N$ if $I_k = 1/2$ $\forall
k$. The profile of the hyperfine bias distribution can be calculated
with the knowledge of the individual couplings, and is well
described by a Gaussian with half-width $E_{0} = \sum_k
[(I_k+1)/3I_k](\omega_k^{\parallel}I_k)^2 \simeq 0.082$
K.\cite{stamp04CP}

In addition to the hyperfine couplings, the $\mathbf{S}$ spins are also mutually coupled by dipolar interactions,
which yield an additional bias $\xi_D = 2g \mu_{\rm B} \mathbf{S}\cdot \mathbf{B}_{\rm dip}$. The dipolar bias
can be considered quasi-static in the sense that it remains essentially constant over time intervals that are
long compared to the typical timescale for the hyperfine bias fluctuations. The distribution of dipolar biases
depends on the total magnetization of the sample and, in general, on its shape. For a demagnetized, ZFC sample of
Mn$_{12}$-ac, the dipolar bias distribution is described by a Gaussian with half-width $E_{\rm D} \simeq 0.32$
K.\cite{tupitsynP} Finally, one may in general apply a static external field, $B_z$, along the $\hat{z}$-axis,
which produces an additional bias $\xi_B = 2g \mu_B S_z B_z$. For zero external field and some typical nonzero
value of $\xi_{D}$, the energy level scheme of a Mn$_{12}$-ac cluster coupled to its nuclear spins would resemble
the sketch shown in Fig. \ref{tunnelbias}.

To analyze the behavior of this system with respect to incoherent
tunneling of the electron spin, the crucial question to be answered
is what happens to the nuclear spins when $\mathbf{S}$ suddenly
changes direction. How many of the $\{\mathbf{I}_k\}$ coflip with
$\mathbf{S}$? As extensively discussed in the PS
literature,\cite{prokof'ev95CM,prokof'ev96JLTP,prokof'ev00RPP} there
are two mechanisms by which nuclear spins may be flipped by a
tunneling event. First, a nuclear spin may coflip with $\mathbf{S}$
if the local hyperfine field does not exactly reverse its direction
after $\mathbf{S}$ has tunnelled, since it would then start to
precess around a different axis, hence the name ``orthogonality
blocking'' or ``precessional decoherence'' for this mechanism. The
number of spins coflipped this way is $\kappa$, defined as (see Fig.
\ref{PSangles} for $\beta_k$):
\begin{subequations}
\begin{eqnarray}
e^{- \kappa} = \prod_k \cos \beta_k \approx e^{-\frac{1}{2}\sum_k \beta_k^2}, \label{defkappa}\\
\cos(2 \beta_k) = \frac{-\mathbf{B}_{{\rm hyp},k}^{(1)} \cdot \mathbf{B}_{{\rm hyp},k}^{(2)}}{|\mathbf{B}_{{\rm
hyp},k}^{(1)}||\mathbf{B}_{{\rm hyp},k}^{(2)}|} .
\end{eqnarray}
\end{subequations}
The cosine factors in Eq.~(\ref{defkappa}), which are multiplied
over all the bath spins, are the overlap matrix elements between the
initial and final bath states, i.e. $\langle i|U_k|f\rangle$, where
$U_k$ is the rotation operator of the $k$-th bath spin (Ref.
\onlinecite{prokof'ev00RPP}, Appendix A.2). Clearly, $\kappa$
depends only on the direction of the hyperfine fields, and not on
the timescale of the electron spin flip. The nuclei in Mn$^{(1)}$
sites do not contribute to $\kappa$ since the $\mathbf{B}_{\rm hyp}$
before and after the flip are exactly antiparallel, i.e.,
$\omega^{\perp}(\mathrm{Mn}^{(1)}) = 0$. Conversely, $^{1}$H nuclei
in the ligands may give a large contribution because they are
subject to the vector sum of the dipolar fields from several
molecules, which does not entirely reverse direction when just one
molecule flips.

The other possibility is that the nuclear spins follow adiabatically
the rotation of $\mathbf{S}$. For this to happen, the ``bounce
frequency'' of $\mathbf{S}$, $\Omega_0$, has to be small or
comparable with the nuclear Larmor frequencies. $\Omega_0$ is given
here by the energy difference between the $|m|=S$ and $|m|=S-1$
cluster spin states: since we are interested in FRMs, knowing that
the resonance between $m=-S$ and $m=S-1$ states occurs at $B_z
\simeq 0.39$ T (Ref. \onlinecite{wernsdorfer99EPL}) yields $\hbar
\Omega_0 \simeq 10$ K, i.e., several orders of magnitude larger than
$\{\omega_k\}$. Therefore, the nuclei cannot adiabatically follow
the dynamics of $\mathbf{S}$ and the number of spins coflipped by
this mechanism, $\lambda$, is essentially zero. As a matter of
nomenclature, this mechanism leads to what is called ``topological
decoherence'' because the topological phase of the
$\{\mathbf{I}_k\}$ becomes entangled with that of
$\mathbf{S}$.\cite{prokof'ev95CM,prokof'ev96JLTP,prokof'ev00RPP}

\begin{figure}[t]
\includegraphics[width=6cm]{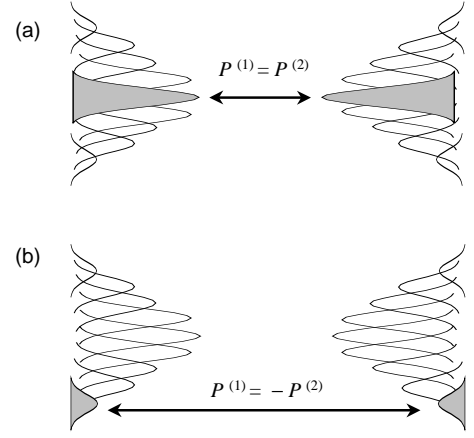}
\caption{\label{coflip} Sketch of two resonant tunneling processes, differing in the number of required nuclear
coflips. (a) The nuclear polarization is the same before and after the electron spin flip: this process has
maximum likelihood. (b) All nuclei need to reverse their spin to conserve the total energy: this process is
extremely unlikely.}
\end{figure}

Combining the two flipping mechanisms defines a parameter $\xi_0
\propto \lambda + \kappa$, which expresses how much the nuclear
polarizations before and after the electron spin flip may differ for
the flip to be likely to occur. Two opposite situations are sketched
in Fig. \ref{coflip}, where we call $\mathcal{P}^{(1)}$ and
$\mathcal{P}^{(2)}$ the nuclear polarizations before and after the
electron spin flip, respectively. In any case the system has to
tunnel between states at the exact resonance, but in case (a) the
electron flip does not require any nuclear coflip
($\mathcal{P}^{(1)}=\mathcal{P}^{(2)}$), while case (b) requires
\emph{all} nuclei to coflip ($\mathcal{P}^{(1)} = -
\mathcal{P}^{(2)}$), which is extremely unlikely. As a result, the
expression for the tunneling rate contains a factor $\exp(-\xi /
\xi_0)$ that describes precisely this restriction. From the above
discussion it is clear that - at least in the absence of external
transverse fields\cite{tupitsyn04PRB} - the main contribution to
$\xi_0$ comes from $^1$H nuclei, and that $\xi_0 \ll
\{E_{0},E_{D}\}$.

In the presence of a dipolar bias, the tunneling transition with highest probability, i.e. no coflipping nuclei,
occurs when $\xi_{D} = \xi_{N}$ (Fig. \ref{tunnelbias}). This means that a tunneling event effectively entails an
exchange of dipolar and hyperfine energy. We may then distinguish between transitions that increase the hyperfine
energy (``left to right'' in Fig. \ref{tunnelbias}), occurring at a rate $w_{\uparrow}$, and transitions that
decrease it (``right to left '' in Fig. \ref{tunnelbias}) at a rate $w_{\downarrow}$. The total tunneling rate is
$\Gamma^{N} = (w_{\uparrow} + w_{\downarrow})/2$. The PS expressions for $\Gamma^{N}$, generalized to the $m$-th
electronic doublet, given a (dipolar) longitudinal bias $\xi$ on the ground doublet, are:\cite{stamp04CP}
\begin{eqnarray}
\Gamma_m^{N}(\xi) & \simeq & \frac{2 \Delta_m^2 G_N^{(m)}}{\sqrt{\pi} \hbar E_{0}^{(m)}}
\exp(-|\xi^{(m)}|/\xi_0), \label{GammaNxi} \\
G_N^{(m)} & = & \exp[-\Delta_m^2 / 2(E_0^{(m)})^2] \nonumber \\
E_0^{(m)} & \simeq & \sqrt{E_0^2 \frac{m^2}{S^2} - \Delta_m^2}, \nonumber \\
\xi^{(m)} & = & \xi \frac{m}{S}, \nonumber
\end{eqnarray}
where $\Delta_m$ is the tunneling matrix element of the $m$-th electron spin doublet, to be calculated by exact
diagonalization of the giant spin Hamiltonian. The factor $G_N^{(m)}$ expresses the fact that the spin-bath
mediated tunneling rate vanishes when $\Delta_m \gg E_0^{(m)}$, i.e. when the spread of nuclear energies is not
sufficient to sweep the hyperfine bias through the tunneling resonance. The parameters $E_0^{(m)},\xi^{(m)}$ are
generalizations to arbitrary electron spin doublets of the quantities $E_0$ and $\xi$ defined before for the
ground doublet $|m|=S$, while $\xi_0$ is assumed $m$-independent. To obtain the total tunneling rate through the
$m$-th doublet we average Eq. (\ref{GammaNxi}) over the distribution of dipolar biases:
\begin{eqnarray}
P_m(\xi) = \frac{1}{\sqrt{2 \pi}E_D^{(m)}} \exp(-\xi^2/2E_D^{(m)}), \label{Pdip}
\end{eqnarray}
In the real situation considered here, the spread of dipolar biases
in the sample is much larger than the tunneling window allowed by
hyperfine couplings, $E_{\rm D} \gg \xi_0$. This means we can
estimate $\Gamma^N_m$ by calculating the fraction $x_m^N$ of
molecules with bias $-\xi_0^{(m)} < \xi < \xi_0^{(m)}$, for which we
may approximate $\Gamma_m^{N}(\xi) \simeq \Gamma_m^{N}(0)$, and by
neglecting the contribution of the molecules whose bias is larger
than $\xi_0$ and which tunnel at an exponentially small rate:
\begin{eqnarray}
\int_{-\infty}^{+\infty} P_m(\xi) \Gamma_m^N (\xi) \mathrm{d}\xi \simeq x_m^N \Gamma_m^N (0), \\
x_m^N = \int_{-\xi_0}^{+\xi_0} P_m(\xi) \mathrm{d}\xi.
\end{eqnarray}

Finally, the global spin-bath driven tunneling rate, $\Gamma^N$, is obtained by summing over the $m$ electronic
doublets weighed with the appropriate Boltzmann occupation factor:
\begin{eqnarray}
\Gamma^N(T) \simeq \frac{1}{Z} \sum_m \exp\left(-\frac{E_m}{k_{\rm B}T}\right) x_m^N \frac{2 \Delta_m^2
G_N^{(m)}}{\sqrt{\pi} \hbar E_{0}^{(m)}}, \label{GammaN}
\end{eqnarray}
where $\{E_m\}$ are the average energies of the $m$-th doublets and $Z$ is the partition function. Notice that
the spin-bath driven tunneling rates are individually $T$-independent: the temperature enters only in the
Boltzmann factors for the occupation of the $m$-th doublets, and thereby in their contribution to the global
tunneling rate $\Gamma^N(T)$. Also, since $P_m(\xi)$ is essentially constant in the interval $-\xi_0 < \xi <
\xi_0$, we have $\Gamma_m^N \propto x_m^N \propto \xi_0$. This immediately explains why the isotopic substitution
of $^{2}$H for $^{1}$H yields a decrease in tunneling rate [Fig. \ref{T1T2D}(a)], since these are the nuclei that
mostly contribute to $\xi_0$.

\subsection{Phonon-induced tunneling rate}
\label{phononrate}

For comparison, we also discuss the case where the electron spin tunneling is caused by spin-phonon couplings.
The phonon-driven tunneling rate though the $m$-th doublet, $\Gamma^{\phi}_m$, is related to the ($T$-dependent)
broadening of the electron spin states, $w_m(T)$, by:\cite{kagan80JETP,stamp04CP}
\begin{eqnarray}
\Gamma_m^{\phi}(\xi) \simeq \frac{\Delta_m^2 w_m(T)}{\xi_m^2 + \Delta_m^2 + \hbar^2 w_m^2(T)} \label{Gammaphi}
\end{eqnarray}
The phonon-induced broadenings are obtained as a function of the sample density, $\rho$, the sound velocity,
$c_s$, and the uniaxial anisotropy parameter, $D$, as:\cite{leuenberger00PRB}
\begin{subequations}
\begin{eqnarray}
w_m(T)  =  p_{m+1,m} + p_{m-1,m} + p_{m+2,m} + p_{m-2,m} ,\\
p_{m\pm 1,m}  =  s_{\pm 1} \frac{D^2}{12 \pi \rho c_s^5 \hbar^4} \frac{(E_{m \pm 1} - E_m)^3}{e^{(E_{m \pm 1} - E_m)/k_{\rm B}T} - 1} , \\
p_{m \pm 2,m}  =  s_{\pm 2} \frac{17 D^2}{192 \pi \rho c_s^5 \hbar^4} \frac{(E_{m \pm 2} - E_m)^3}{e^{(E_{m \pm
2} - E_m)/k_{\rm B}T} - 1} ,
\end{eqnarray}
\end{subequations}
with $s_{\pm 1} =  (S \mp m)(S \pm m+1)(2m \pm 1)^2$ and $s_{\pm 2}  =  (S \mp m)(S \pm m+1)(S \mp m -1)(S \pm
m+2)$.

Again, we calculate the fraction of molecules with highest tunneling rate, $x_m^{\phi}$, as those whose bias is
within the width of the Lorentzian function (\ref{Gammaphi}):
\begin{eqnarray}
x_m^{\phi} = \int_{-\sqrt{\Delta_m^2 + \hbar^2 w_m^2(T)}}^{+\sqrt{\Delta_m^2 + \hbar^2 w_m^2(T)}} P_m(\xi)
\mathrm{d}\xi,
\end{eqnarray}
and weigh the contribution of the $m$-th levels with their Boltzmann factor to obtain the total phonon-driven
tunneling rate:
\begin{eqnarray}
\Gamma^{\phi}(T) \simeq \frac{1}{Z} \sum_m \exp\left(-\frac{E_m}{k_{\rm B}T}\right) x_m^{\phi} \frac{\Delta_m^2
w_m(T)}{\Delta_m^2 + \hbar^2 w_m^2(T)}. \label{GammaP}
\end{eqnarray}
Contrary to the nuclear-driven case, here each individual $m$-th doublet tunneling rate $\Gamma_m^{\phi}$ is
$T$-dependent by itself, besides being weighed by Boltzmann factors. This means that the phonon-driven tunneling
rate never shows a $T$-independent plateau, even at very low-$T$ when tunneling occurs only through the ground
doublet.

\subsection{Tunneling rate - longitudinal spin relaxation rate}

To relate the electron spin tunneling rates, $\Gamma^{N,\phi}(T)$, to the observed LSR rate, $W$, we apply the
remarks made above on the behavior of the nuclear spins upon a sudden change of the electron spin direction. In
particular, we shall compare theory and experiments for the Mn$^{(1)}$ site, where (ideally) the hyperfine
coupling would be strictly scalar.\cite{kubo02PRB} This implies that the $^{55}$Mn spin at Mn$^{(1)}$ sites do
not coflip by ``precessional decoherence'', since $\omega_k^{\perp} \simeq 0$, neither do they coflip by
``topological decoherence'' due to the very small values of $\omega_k^{\parallel}/\Omega_0$. Thus, each electron
spin tunneling event corresponds to the inversion of the populations of the \emph{local} nuclear Zeeman levels.
The LSR rate arising from this situation is easily obtained and can be found in the
literature,\cite{abragam61,alexander65PR,morelloT,baek05PRB} but we repeat here the derivation because it will
allow us to point out exactly why no known theory can explain our data. The answer is most easily obtained for
nuclear spins $I=1/2$, but remains valid for arbitrary spin values.

We start by writing a master equation for the populations of the nuclear Zeeman levels \emph{relative to the
local hyperfine field direction}, calling $N_+$ the number of nuclei in the excited Zeeman state and $N_-$ those
in the ground state. For simplicity, since the internal equilibrium is reestablished within a time $T_2 \ll
\tau_{\rm T}$ after each tunneling event, we assume that just before tunneling all clusters have the same values
of $N_+$ and $N_-$, neglecting fluctuations around the mean values. If a fast-relaxing molecule tunnels at time
$t$, the polarization of its own nuclei is abruptly inverted. Each time a tunneling transition lowers the energy
of the local nuclei, which occurs at a rate $w_{\downarrow}$, then $N_-$ nuclei have been added to the total
number of nuclei in the Zeeman ground state. After a time $T_2$ this decrease in local hyperfine bias has been
redistributed over the sample: calling $x_{\rm FRM}$ the fraction of FRMs over the total, then the tunneling
event has increased $N_+$ to $N_+ + x_{\rm FRM}N_-$. The same reasoning holds for transitions that increase the
hyperfine bias. The master equation is therefore:
\begin{eqnarray}
\frac{\mathrm{d}N_+}{\mathrm{d}t} = x_{\rm FRM}N_- w_{\downarrow} - x_{\rm FRM}N_+ w_{\uparrow}.
\label{mastereq}
\end{eqnarray}
From here the LSR rate can be obtained by standard textbook calculations.\cite{slichterB} Writing $N_+=(N+n)/2$
and $N_-=(N-n)/2$, Eq. (\ref{mastereq}) becomes:
\begin{eqnarray}
\frac{\mathrm{d}n}{\mathrm{d}t} = x_{\rm FRM}N(w_{\downarrow} - w_{\uparrow}) - x_{\rm FRM}n(w_{\downarrow} +
w_{\uparrow}),
\end{eqnarray}
which can be rewritten as:
\begin{eqnarray}
\frac{\mathrm{d}n}{\mathrm{d}t} = 2W(n_0 - n),\label{masterfinal}\\
n_0 = N \frac{w_{\downarrow} - w_{\uparrow}}{w_{\downarrow} +
w_{\uparrow}},\\
W = x_{\rm FRM} \frac{w_{\downarrow} + w_{\uparrow}}{2} = x_{\rm FRM} \Gamma^{N,\phi} \label{Wmodel},
\end{eqnarray}
where $n_0$ is the equilibrium nuclear polarization and $W$ is the desired LSR rate, since the solution of
(\ref{masterfinal}) is precisely of the form $n(t) = n(0) - [n(0)-n_0][1-\exp(-2Wt)]$.

\begin{figure*}[t]
\includegraphics[width=11cm]{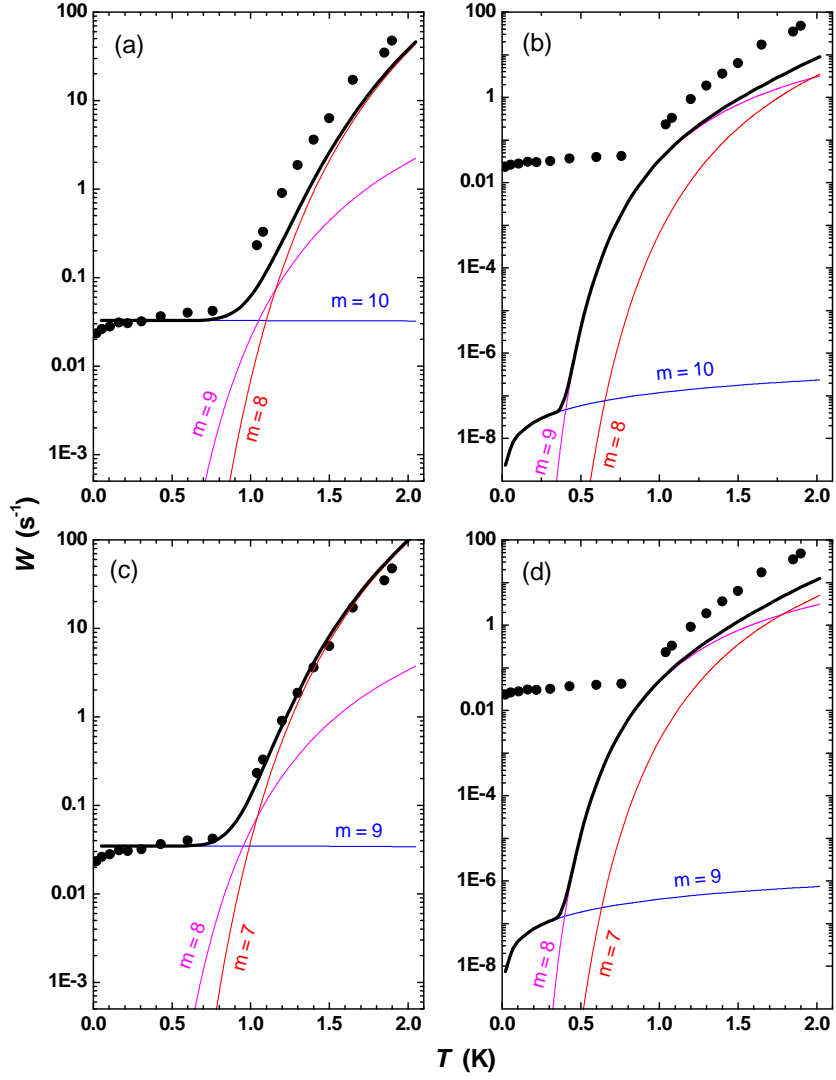}
\caption{\label{fitW1}(Color online) Calculated nuclear spin-lattice
relaxation rates, $W$, as a function of temperature, for spin-bath
[panels (a) and (c)] and phonon [panels (b) and (d)] mediated
tunneling. In panels (a) and (b) a total spin $S=10$ is assumed for
the FRMs; in panels (c) and (d), $S=9$. The complete parameter sets
are given in table \ref{tablefitW1}. Black dots: experimental data;
thick black lines: calculated $W$; thin lines: contributions of the
$m$-th electron spin doublets to the total rate.}
\end{figure*}

We now attempt to fit the measured LSR rate in zero field at the Mn$^{(1)}$ site in ZFC sample, using Eqs.
(\ref{GammaN}), (\ref{Gammaphi}) and (\ref{Wmodel}). To this end we calculate the energy levels scheme of the
FRMs using the effective spin Hamiltonian:
\begin{eqnarray}
\mathcal{H}_{\rm FRM} = -DS_z^2 + E(S_x^2 - S_y^2) - C(S_+^4 + S_-^4). \label{Hfrm}
\end{eqnarray}
Unfortunately, very little is known about the parameter values in (\ref{Hfrm}). To the best of our knowledge,
it's not even established whether FRMs in Mn$_{12}$-ac have lowest total spin state $S=10$ or, for instance,
$S=9$. The analysis of a Mn$_{12}$ variant containing only FRMs\cite{takeda02PRB} seemed to support an $S=10$
ground state, but it's not clear to what extent the FRMs in Mn$_{12}$-ac have the same properties as those
analyzed in Ref. \onlinecite{takeda02PRB}. For instance, Ref. \onlinecite{takeda02PRB} finds the first level
crossing transition, i.e. the value of longitudinal field at which $E_{S} = E_{-S+1}$, at $B_z^{S,-S+1} \simeq
0.27$ T, quite different from the value observed for the actual FRMs in Mn$_{12}$-ac, $B_z^{S,-S+1} \simeq 0.39$
T.\cite{wernsdorfer99EPL} We shall try both $S=10$ and $S=9$ and discuss how the different behaviors compare to
the experimental data. To avoid having too many fitting parameters, we choose to keep $C$ fixed at the value
commonly used for the majority species of Mn$_{12}$-ac, $C = 4.4 \times 10^{-5}$ K.\cite{mirebeau99PRL} The
uniaxial anisotropy, $D$, is obtained by imposing the condition $B_z^{S,-S+1} \simeq 0.39$
T,\cite{wernsdorfer99EPL} yielding $D = g \mu_{\rm B} \times 0.39 = 0.524$ K. Although we tried adding also a
fourth order term, $-BS_z^4$, it turned out that the best fits are obtained by leaving $B=0$, so we shall not
discuss this further. The rhombic anisotropy term, $E$, is used as the actual fitting parameter since it most
directly influences the value of the tunneling splittings $2\Delta_m$ and thereby the tunneling rates
$\Gamma^{N,\phi}$. Carretta \textit{et al.}\cite{carretta04PRL} showed that the effective $\Delta$ is extremely
sensitive to the gap between the lowest lying total spin manifolds (``S-mixing''), and could explain why the
observed Landau-Zener tunneling probabilities in Fe$_8$ are much larger than what would be expected on basis of
the spin Hamiltonian parameters for the $S=10$ manifold.\cite{wernsdorfer99S} A small gap between lowest lying
total spin manifolds is quite expectable for FRMs, so the values of $E$ we need to justify the $\Gamma^{N,\phi}$
extracted from experiment should not be taken literally as an estimate of the anisotropy parameter. In other
words, the values of $E$ used in our calculations account also for the possible S-mixing due to an energetically
close manifold with different total spin $S$, and do not necessarily correspond to the values that one would
obtain from neutrons scattering or EPR experiments. We take as fixed parameters the sound velocity $c_s = 1.5
\times 10^3$ m/s (Ref. \onlinecite{mettes01PRB}), the density $\rho = 1.83 \times 10^3$ g/m$^3$ (Ref.
\onlinecite{lis80AC}), $E_{D} = 0.32$ K (Ref. \onlinecite{tupitsynP}) and $E_{0} = 0.082$ K (Ref.
\onlinecite{stamp04CP}), whereas $\xi_0$ is allowed to vary. When comparing nuclear- and phonon- driven tunneling
rates, we impose the same parameters for the spin Hamiltonian (\ref{Hfrm}). The results of the calculations are
shown in Fig. \ref{fitW1}, for the set of parameters given in Table \ref{tablefitW1}.

\begin{table}
\caption{\label{tablefitW1}Parameter values for the calculations shown in Fig. \ref{fitW1}. In bold are given the
free fitting parameters. The tunneling matrix element in the lowest doublet, $\Delta_{S}$, is obtained from the
Hamiltonian (\ref{Hfrm}), i.e. is not introduced by hand.}
\begin{ruledtabular}
\begin{tabular}{cccccccc}
Panel & $S$ & $D$ & $E$ & $\xi_0$ & $\Delta_{S}$ \\
 &  & (K) & (K) & (mK) & ($\mu$K) \\
\hline
a & 10 & 0.524 & \textbf{0.204} & \textbf{10} & 5.4 \\
b & 10 & 0.524 & 0.204 & - &  5.4 \\
c & 9 & 0.524 & \textbf{0.178} & \textbf{10} & 5.6 \\
d & 9 & 0.524 & 0.178 & - & 5.6
\end{tabular}
\end{ruledtabular}
\end{table}

By looking at the theoretical curves alone we find that, given a
fixed set of parameters for the FRMs' spin Hamiltonian (\ref{Hfrm}),
the nuclear-driven tunneling process always dominates over the
phonon-driven one, both in the low-$T$ and in the high-$T$ regime.
The situation may be reversed in the high-$T$ regime by assuming the
sound velocity is lower than the literature value used here, but
$\Gamma^{\phi} > \Gamma^N$ would never hold at low-$T$ under
realistic circumstances. A comparison with the experimental data
shows that both nuclear- and phonon- driven mechanisms yield a
correct slope of $W(T)$ in the thermally-assisted regime, $T>0.8$ K,
whereas the phonon process can never reproduce the low-$T$ plateau.
An almost perfect fit of the data is obtained by assuming that the
FRMs have a total spin $S=9$, while the $S=10$ case has a
$T$-dependent region systematically starting at too high
temperatures. In the nuclear-driven case, we used an optimal value
of $\xi_0 \simeq 10$ mK which seems very reasonable, since the main
contribution to $\xi_0$ arises from the coupling to protons.
Conversely, the values used for $E$ appear very high, since to fit
the NMR data we need to assume $D/E \simeq 2.5$. As mentioned
before, however, such a high value of $E$ should not be interpreted
as the spectroscopic rhombic term in the spin Hamiltonian, since it
is used here as the parameter that tunes the tunneling splitting of
the ground doublet, and therefore incorporates the effect of
$S$-mixing\cite{carretta04PRL} if a manifold with different total
spin value is energetically close to the $S$ ground state.

An important remark is that our calculations, which only account for
tunneling fluctuations as a source of nuclear LSR, can accurately
reproduce the observed LSR rate \emph{also in the thermally
activated regime}, $T > 0.8$ K. All the previous NMR experiments in
that
regime\cite{lascialfari98PRL,furukawa01PRB,goto03PRB,chakov06JACS}
have been interpreted in terms of the ``intrawell'' electron spin
fluctuations, that arise from thermal excitation of the electron
spin state on the same side of the anisotropy barrier [see Fig.
\ref{structure}(b)]. As pointed out by Goto \textit{et
al.},\cite{goto03PRB} that reasoning is inappropriate when applied
to the LSR of nuclei belonging to Mn$^{4+}$ ions, since the
hyperfine coupling tensor is diagonal. In that case, a fluctuation
in the $\hat{z}$ projection of the electron spin does not result in
a fluctuating field perpendicular to the nuclear quantization axis
(which is $\hat{z}$ itself), and cannot account for longitudinal
nuclear spin relaxation. As we show in Fig. \ref{fitW1}, including
the effect of electron spin tunneling through $|m| < S$ doublets
solves what appeared to be a paradox, since the tunneling
fluctuations induce nuclear LSR also in the absence of non-diagonal
hyperfine coupling terms. For the nuclei in Mn$^{3+}$ ions, the
intrawell electron spin transitions do provide an additional channel
for nuclear relaxation, which may explain why the LSR rate at $T >
1$ K in the Mn$^{3+}$ sites is larger than that in the Mn$^{4+}$
ones.\cite{furukawa01PRB,goto03PRB}

\subsection{Thermal equilibrium}
\label{thermaleq}

In the preceding discussion, it may seem that we have not explicitly used the condition that the nuclear spins
are in thermal equilibrium, which is what we observe in the experiment. This condition, however, is automatically
implied in the application of Eq. (\ref{Wmodel}) to the LSR rate. For Eq. (\ref{Wmodel}) to actually represent
the rate at which the nuclear spins exchange energy with a thermal bath and thereby return to the equilibrium
magnetization after a perturbing NMR pulse, one needs to include the detailed balance condition:
\begin{eqnarray}
\frac{w_{\uparrow}}{w_{\downarrow}} = \exp\left(-\frac{\hbar \omega_N}{k_{\rm B}T}\right). \label{detbal}
\end{eqnarray}
In other words, if the nuclear spin temperature has to reach
equilibrium with the thermal (phonon) bath via the process described
by the rates $w_{\uparrow}$ and $w_{\downarrow}$, the latter
\emph{must} satisfy (\ref{detbal}). The detailed balance condition
is often taken for granted, but in this case one needs to be more
careful. $w_{\uparrow}$ and $w_{\downarrow}$ represent the rates of
\emph{electron spin transitions} that increase or lower the nuclear
energy, respectively. The crucial point is that both are rates for
\emph{tunneling} transitions, which occur when the \emph{total
energy} of the ``electron plus nuclear spins'' system is the same
before and after the electron spin flip. Thus, the difference
between $w_{\uparrow}$ and $w_{\downarrow}$ is simply that, e.g.,
$w_{\uparrow}$ is the rate for a tunneling transition that increases
the nuclear spin energy while reducing the electronic one. That is,
after the flip most of the nuclei are oriented against their local
hyperfine field, while the electron spin is favorably aligned with
respect to the local field (in particular the dipolar one, when $B_z
= 0$). $w_{\downarrow}$ does the opposite and, interestingly, this
means that the instantaneous local spin temperature (local referring
to the nuclei belonging to a specific molecule that has just
flipped) is negative. This situation is clearly very different from
the standard NMR picture of nuclear relaxation by coupling to
paramagnetic centers, where the latter make spin-phonon transitions
between Zeeman-split levels having different thermal populations.

Now we can summarize the meaning of our experimental results for the
description of electron spin tunneling in the presence of a nuclear
spin bath:

(i) The Prokof'ev-Stamp theory of the spin bath, as developed so far
and reviewed in \S\ref{spinbath}, quantitatively and qualitatively
reproduces the nuclear LSR rate in the whole temperature regime of
our measurements, by assuming that the LSR is due to tunneling
events in a minority of fast-tunneling molecules;

(ii) The additional observation that the nuclear spins are in
thermal equilibrium with the phonon lattice at all temperatures
implies that \emph{the rates $w_{\uparrow}$ and $w_{\downarrow}$
must be different.} For this to happens, it is necessary to
explicitly include the role of spin-phonon interactions in the
nuclear-spin mediated tunneling process. Importantly, the results of
the calculations shown in Fig. \ref{fitW1}(b,d) indicate that it is
not sufficient to attribute the thermal relaxation to a
phonon-assisted tunneling process as described in
\S\ref{phononrate}, working ``in parallel'' to the nuclear-spin
mediated tunneling process. At the lowest temperatures, even the
longest thermalization times observed in our experiments
(\S\ref{timespinT}) are still much shorter than
$(\Gamma^{\phi})^{-1}$ as calculated from phonon-assisted tunneling
alone (\S\ref{phononrate}), thus reinforcing the need for a theory
that includes nuclear-spin and phonon mediated tunneling \emph{at
the same time};

(iii) Our statement that the nuclear relaxation has to be mediated
by inelastic electron spin tunneling processes, is further supported
by specific heat experiments that show that a system of
dipolarly-coupled tunneling molecules can relax to the long-range
ferromagnetically ordered state, provided that the tunneling rate is
fast enough for the experimental detection of the ordering
anomaly.\cite{evangelisti04PRL} This means that \emph{there is a
mechanism for the ensemble of electron spins to find its
thermodynamic ground state, even at temperatures so low that the
relaxation can only proceed by quantum tunneling. We argue here that
such inelastic tunneling mechanism is the same mechanism that is
responsible for the thermalization of the nuclear spins.} Indeed, by
extending the specific heat measurements below the ordering
temperature, one even observes the equilibrium specific heat
contribution of the nuclear spins, meaning that both the electron
spins and the nuclear spins can attain thermal equilibrium within
the time-scale of the specific heat experiment (10 - 100 s) at all
temperatures reached.

All the work presented here is dedicated to the small-$\Delta_0$,
incoherent tunneling regime for the central spin. Having shown that
the description of the nuclear-spin mediated tunneling is incomplete
without the inclusion of spin-phonon couplings, some concerns may be
raised also on the current description of the spin bath effects on
the electron spin in the large-$\Delta_0$, coherent tunneling
regime.\cite{stamp04PRB,morello06PRL} This work cannot address that
issue, and we think that the answer will have to come from low-$T$
NMR experiments in large transverse field, and \emph{quantitative}
analysis of pulsed-ESR experiments.

\section{CONCLUDING REMARKS} \label{conclusions}

The purpose of the research presented here is to illustrate and analyze a prototypical example of quantum
tunneling of a macroscopic variable (the giant spin of single-molecule magnet) in the presence of a spin bath
environment. Instead of looking at the macroscopic variable itself and deducing the effect of the environment on
its dynamics, as is most often done, we have directly observed the behavior of the spin bath by means of low-T
NMR experiments.

We have provided compelling evidence that the longitudinal nuclear
spin relaxation in the $^{55}$Mn nuclei of Mn$_{12}$-ac is driven by
electron-spin quantum tunneling fluctuations. The nuclear LSR rate,
$W$, indeed contains all the features that are expected to be
associated with tunneling of the molecular spin: i) A
$T$-independent plateau of the LSR rate for $T<0.8$ K; ii) A strong
dependence of $W$ on a longitudinal magnetic field, that destroys
the resonance condition for electron spin tunneling; iii) The
slowing down of the nuclear LSR upon isotropic substitution of $^1$H
by $^2$H in the ligands, by an amount identical to the slowing down
of the quantum relaxation of the magnetization observed in similar
systems. Because of the short timescale of the observed LSR, we
argued that the tunneling fluctuations must take place in a minority
of fast-relaxing molecules, which are indeed known to be present in
Mn$_{12}$-ac. For these fluctuations to relax the nuclear
magnetization in the entire sample, an additional mechanism is
required which equilibrates the nuclear spin polarization across
neighboring molecules, i.e. intercluster nuclear spin diffusion. Our
data on the transverse nuclear spin relaxation show that the
intercluster spin diffusion is indeed present and effective. All the
above observations confirm and support the picture of nuclear-driven
quantum tunneling of magnetization as originally formulated by
Prokof'ev and Stamp. However, a crucial outcome of our experiments
is the demonstration that the nuclear spins are in thermal
equilibrium with the lattice phonons down to the lowest
temperatures, where only quantum tunneling fluctuations of the
electron spins are still present. This observation cannot be
explained within the present theory of the spin bath.

The implications of our results are potentially very profound,
particularly because of the growing interest toward a coherent
manipulation of spins for quantum information processing. The
spin-bath environment, describing localized two-level systems, has
been repeatedly identified as the most important source of
decoherence in solid-state qubits. This includes superconducting
systems,\cite{martinis05PRL} quantum
dots,\cite{koppens05S,johnson05N} NV centers in
diamond\cite{childress06S} and, of course, molecular
magnets.\cite{ardavan07PRL} We have investigated here the incoherent
tunneling regime, but the theoretical formalism to describe the
coupling between central spin and spin bath is identical in the case
of coherent spin dynamics. Therefore, the main finding of our work -
that the role of phonons in the nuclear-spin mediated tunneling is
currently lacking a proper description - suggests that also the
contribution of the nuclear spin bath to the decoherence rate of
realistic spin qubits may need to be revisited.

\begin{acknowledgments}
We are indebted to O. N. Bakharev, H. B. Brom, D. Bono, N. J.
Zelders and G. Frossati for experimental help and extensive
discussions. Continuous and illuminating theoretical support from P.
C. E. Stamp and I. S. Tupitsyn is gratefully acknowledged, and so
are discussions with W. Wernsdorfer, S. Hill, N. V. Prokof'ev, B. V.
Fine, M. Evangelisti, A. J. Leggett, Y. Imry, M. Schechter and A. L.
Burin. We also thank K. Awaga and K. Takeda for useful
correspondence about their results in Ref. \onlinecite{takeda02PRB}.
The Mn$_{12}$-ac samples were supplied by R. Sessoli and A. Caneschi
(crystallites, natural and deuterated) and A. Millan (single crystal).\\
This work is part of the research program of the ``Stichting FOM'' and is partially funded by the EC-RTN
``QuEMolNa'' and EC-Network of Excellence ``MAGMANet'' (No.515767-2).
\end{acknowledgments}

%\bibliographystyle{unsrt}
 %\bibliography{Mn12PRB}

\begin{thebibliography}{10}
\bibitem{noteanisotropy}
{The anisotropy parameters obtained by EPR seem to depend on the
magnetic field
  range used in the experiment \cite{barra97PRB,hill98PRL}, whereas neutron
  scattering is a zero-field experiment, and does not require assumptions on
  the $\mathbf{g}$ tensor in order to fit the data.}
\end{thebibliography}

\end{document}